\documentclass[aps, prd, tightenlines, letterpaper, amsmath, amssymb, preprintnumbers, floatfix, longbibliography, nofootinbib, twocolumn]{revtex4-2}

\usepackage{latexsym,amsfonts,bm}
\usepackage{xspace}
\usepackage{cancel}
\usepackage{graphicx}    
\usepackage{microtype}   
\usepackage{amsmath}     
\usepackage{amsthm}    
\usepackage{amssymb}   
\usepackage{siunitx}     
\usepackage{booktabs}    
\usepackage{hyperref}
\hypersetup{colorlinks=true,citecolor=blue,linkcolor=blue,breaklinks=true}
\usepackage[capitalize]{cleveref}
\usepackage{tensor}
\usepackage[normalem]{ulem}
\usepackage{orcidlink}

\usepackage{adjustbox}    
\usepackage{subcaption}
\usepackage{mathtools} 
\usepackage{bm}
\usepackage{dsfont}
\usepackage[shortlabels]{enumitem}  
\usepackage{physics}   

\allowdisplaybreaks

\providecommand{\CE}{\mathcal{E}}

\providecommand{\CN}{\mathcal{N}}

\providecommand{\CP}{\mathcal{P}}

\providecommand{\nn}{\nonumber}
\providecommand{\uone}{$\mathrm{U}(1)$\xspace}
\providecommand{\sutwo}{$\mathrm{SU}(2)$\xspace}
\providecommand{\suthree}{$\mathrm{SU}(3)$\xspace}

\providecommand{\sutwo}{$\mathrm{SU}(2)$\xspace}

\newcommand{\htp}{H_{\square\square}}
\newcommand{\hathtp}{\hat{H}_{\square\square}}

\DeclareMathOperator*{\argmin}{arg\,min}
\usepackage{accents}
\usepackage{booktabs}
\usepackage{tabularx}
\usepackage{graphicx}
\usepackage{multirow}
\usepackage{adjustbox} 

\newcommand{\SubFigRef}[2]{\ref{#1}{\color{blue}{#2}}}

\begin{document}

\begin{figure}
  \vskip -1.cm
  \leftline{\includegraphics[width=0.15\textwidth]{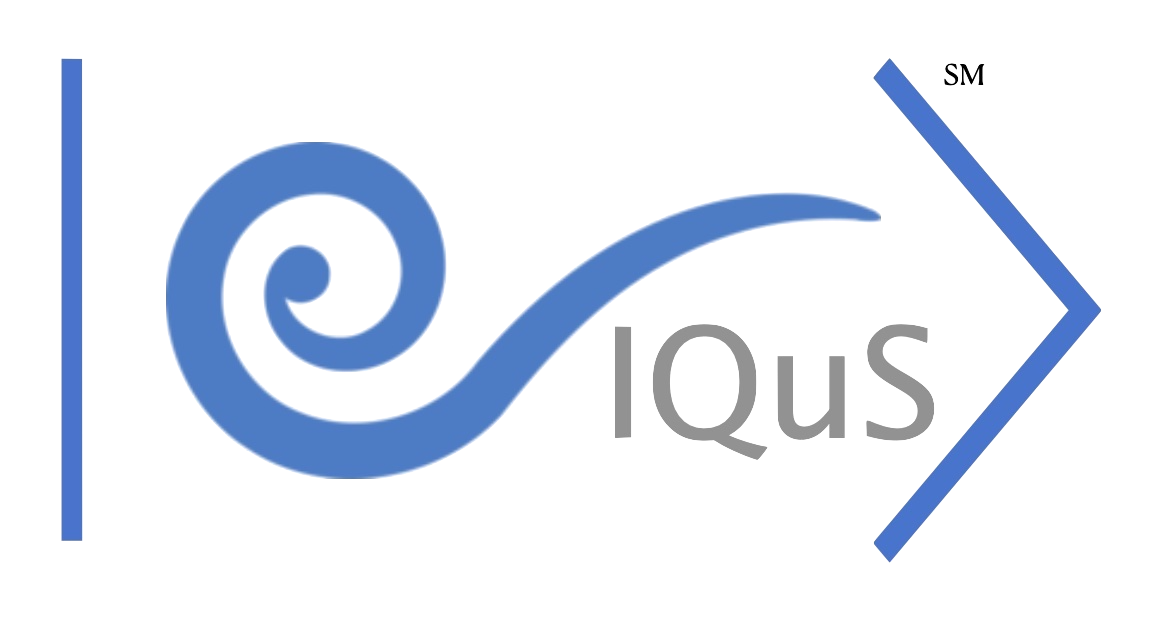}}
\end{figure}

\title{Simulating Fully Gauge-Fixed SU(2) Hamiltonian Dynamics \\ on Digital Quantum Computers}

\author{Henry Froland\,\orcidlink{0009-0008-4356-0602}}
\email[Corresponding author:]{frolandh@uw.edu}
\affiliation{InQubator for Quantum Simulation (IQuS), Department of Physics, University of Washington, Seattle, WA 98195}

\author{Dorota M. Grabowska\,\orcidlink{0000-0002-0760-4734}}
\email{grabow@uw.edu}
\affiliation{InQubator for Quantum Simulation (IQuS), Department of Physics, University of Washington, Seattle, WA 98195}

\author{Zhiyao Li\,\orcidlink{0000-0002-7614-8496}}
\email{zhiyaol@uw.edu}
\affiliation{InQubator for Quantum Simulation (IQuS), Department of Physics, University of Washington, Seattle, WA 98195}

\preprint{IQuS@UW-21-117}
\date{\today}

\begin{abstract}
Quantum simulations of many-body systems offer novel methods for probing the dynamics of the Standard Model and its constituent gauge theories. Extracting low-energy predictions from such simulations rely on formulating systematically-improvable representations of lattice gauge theory Hamiltonians that are efficient at all values of the gauge coupling. One such candidate representation for SU(2) is the fully gauge-fixed Hamiltonian defined in the mixed basis. This work focuses on the quantum simulation of the smallest non-trivial system: two plaquettes with open boundary conditions. A mapping of the continuous gauge field degrees of freedom to qubit-based representations is developed. It is found that as few as three qubits per plaquette is sufficient to reach per-mille level precision on predictions for observables. Two distinct algorithms for implementing time evolution in the mixed basis are developed and analyzed in terms of quantum resource estimates. One algorithm has favorable scaling in circuit depth for large numbers of qubits, while the other is more practical when qubit count is limited. The latter algorithm is used in the measurement of a real-time observable on IBM’s Heron superconducting quantum processor, $\textbf{ibm\_fez}$. The quantum results match classical predictions at the percent-level. This work lays out a path forward for two- and three-dimensional simulations of larger systems, as well as demonstrating the viability of mixed-basis formulations for studying the properties of SU(2) gauge theories at all values of the gauge coupling.

\end{abstract}

\maketitle

\newpage
\tableofcontents

\section{Introduction}
Gauge theories describe a wide array of physical systems from condensed matter and nuclear physics~\cite{PHENIX:2004vcz,BRAHMS:2004adc,PHOBOS:2004zne,STAR:2005gfr,He:2022ywp} to the fundamental interactions of the Standard Model~\cite{GLASHOW1961579,PhysRevLett.19.1264,Salam:1968rm,PhysRevLett.30.1343,politzer1973,higgs1964,Majorana1937SymmetricElectron}. A compelling example is Quantum Chromodynamics (QCD), an \suthree gauge theory which provides a precise and quantitative description of the strong nuclear force over a broad range of energies. Due to its strongly-coupled nature at low energies, it gives rises to a complex array of emergent phenomena that cannot be identified from the underlying degrees of freedom. Additionally, \textit{ab-initio} calculations of its properties are crucial for comparing theoretical predictions of the Standard Model to experimental results. A fully-systematic approach to carry out computations of QCD in the low-energy regime is numerical estimation of the lattice regulated path integral via Monte Carlo importance sampling. This approach is computationally expensive, requiring high-performance computing, but has been remarkably successful in computing certain properties of QCD.\footnote{For an in-depth review, see Ref.~\cite{ParticleDataGroup:2024cfk}} For example, many single-hadron observables, such as the hadron vacuum polarization contribution to the muon anomalous magnetic moment and the hadron spectrum with QED and isospin breaking effects, can be calculated with sub-percent accuracy ~\cite{Blum:2002ii}. Additionally, several two-hadron observables, such as the decay amplitude  for $K \rightarrow \pi \pi$ and direct CP violation can be reliably extracted from lattice QCD simulations~\cite{RBC:2015gro}. 
However, these classical simulation methods encounter severe sign problems when applied to real-time dynamics or systems with finite chemical potential. 

In recent years, there has been growing interest in utilizing quantum computers to study physical processes that appear out of reach to classical approaches. Quantum simulation provides a powerful alternative framework for exploring large many-body quantum systems, enabling first-principle investigations of out-of-equilibrium dynamics~\cite{Benioff1980QuantumTM,Feynman1982Simulating,Feynman1986QuantumComputers,Lloyd1996UniversalQS,schachenmayer2015thermalization,Nandkishore:2014kca,doi:10.1126/science.aaf6725,daley2022practical,mueller2025quantum,cochran2024visualizing,gonzalez2025observation,Bauer:2019qxa,Halimeh:2025vvp,Kim2023UtilityQC,Dumitrescu:2021uin,Hayata:2024smx,Angelides:2023noe,Meth:2023wzd,Kokail:2018eiw,Yang:2020yer,Mil:2019pbt,Bauer:2021gup,Mark:2025wuo,Zhou:2021kdl,Gyawali:2024hrz,Paulson:2020zjd,Turro:2024pxu} and the discovery of novel phases of quantum matter~\cite{Kandala:2017vok,Scholl:2020hzx,Semeghini:2021wls,doi:10.1126/science.abi8378,Thompson:2023kxz,Abanin:2025rbz,Morvan:2023inh,Shinjo:2024vci,Mueller:2022xbg,Than:2024zaj,Schuster:2023klj}.  These methods have already been applied to a variety of gauge groups including $\mathbb{Z}_2$~\cite{Mildenberger:2022jqr,Pardo:2022hrp}, \uone~\cite{Banerjee:2012pg,Hauke:2013jga,Zohar:2013zla,Kuhn:2014rha,Kasper:2015cca,Zohar:2015hwa,Martinez:2016yna,Yang:2016hjn,Kokail:2018eiw,Klco:2018kyo,Lu:2018pjk,Kaplan:2018vnj,Mil:2019pbt,Davoudi:2019bhy,Surace:2019dtp,Haase:2020kaj,Luo:2019vmi,Shaw:2020udc,Yang:2020yer,Ott:2020ycj,Paulson:2020zjd,Nguyen:2021hyk,Zhou:2021kdl,Riechert:2021ink,Bauer:2021gek,Grabowska:2022uos,Kane:2022ejm,zhang2023observation,Farrell:2023fgd,Nagano:2023uaq}, \sutwo~\cite{Zohar:2012xf,Stannigel:2013zka,Mezzacapo:2015bra,Mathur:2015wba,Raychowdhury:2018osk,Raychowdhury:2019iki,Klco:2019evd,Dasgupta:2020itb,Davoudi:2020yln,Atas:2021ext,ARahman:2021ktn,Osborne:2022jxq,halimeh2022gauge,ARahman:2022tkr,zache2023quantum,Alexandru:2023qzd,DAndrea:2023qnr,Fontana:2024rux,Turro:2024pxu,Grabowska:2024emw}, and \suthree~\cite{Anishetty:2009nh,Alexandru:2019nsa,Ciavarella:2021nmj,Farrell:2022wyt,Farrell:2022vyh,Atas:2022dqm,Ciavarella:2021lel,Ciavarella:2023mfc,hayata2023qdeformedformulationhamiltonian,Farrell:2024fit,Ciavarella:2024fzw, Balaji:2025afl}. Remarkable progress on both the algorithmic~\cite{Jordan:2012xnu,Klco:2020aud,Carena:2022kpg,De:2024smi,Gu:2021hyo} and hardware~\cite{IBMQuantumSummit2023,QuantinuumRoadmap2024,Chiu:2025uis,Bluvstein:2023zmt,Evered:2023wbk,Finkelstein:2024zgp} fronts has pushed the frontier of near-term quantum simulations to systems exceeding 100 qubits~\cite{Zemlevskiy:2024vxt,Yu:2022ivm,Shtanko:2023tjn,Baumer:2023vrf,Chen:2023tfg}. These advances have opened the door to studying a broad range of phenomena in strongly interacting theories, such as scattering in high-energy collisions~\cite{Zemlevskiy:2024vxt, Farrell:2025nkx,Davoudi:2024wyv,Chai:2023qpq,Schuhmacher:2025ehh,Bauer:2025nzf,Illa:2022jqb,Nguyen:2021hyk,Briceno:2023xcm}, processes in nuclear physics~\cite{Illa:2022zgu,Amitrano:2022yyn,zhang2023observation,Baroni:2021xtl,Chernyshev:2025lil,Farrell:2022vyh,Farrell:2023fgd,Siwach:2023wzy,Kavaki:2024ijd}, string-breaking~\cite{Gonzalez-Cuadra:2024xul,de2024observation,Alexandrou:2025vaj,cochran2024visualizing,Ciavarella:2024lsp} and hadronization and energy loss~\cite{Farrell:2024mgu, Li:2025sgo, Barata:2025hgx}. 

A key step toward realizing large-scale lattice gauge theory simulations on quantum hardware is the development of efficient, physically-faithful representations of the formally infinite Hilbert space. The fundamental task is encoding the continuous gauge group into discrete degrees of freedom that can be implemented on a quantum device. There exist a variety of approaches that utilize group-theoretic constructions to address this task, including group irreducible representation (irrep) bases~\cite{Mathur:2004kr, mathur2006loop, mathur2007loop, Raychowdhury:2018osk, Raychowdhury:2019iki, Kadam:2022ipf, Kadam:2024zkj, Ciavarella:2021nmj, Klco:2019evd, Davoudi:2020yln, Byrnes:2005qx, Anishetty:2009nh, Banuls:2017ena}, group-element bases~\cite{Mathur:2015wba, DAndrea:2023qnr, Romiti:2023hbd,Grabowska:2024emw}, discrete subgroups~\cite{Alam:2021uuq, Alexandru:2021jpm, Gustafson:2022xdt, Gustafson:2023kvd, Gustafson:2024kym, Assi:2024pdn, Lamm:2024jnl, Muarari:2024dqx}, deformed gauge groups~\cite{Zache:2023dko,Hayata:2023bgh}, and dual bases~\cite{Kaplan:2018vnj, Bender:2020ztu, Haase:2020kaj,Miranda-Riaza:2025fus}. Each of these choices, however, presents its own particular limitations, ranging from inefficiency at weak coupling, complications with systematic improvement, and questions about universality classes as the gauge group is taken to be continuous. Therefore, it is of vital importance to compare these various formulations, with a particular focus on the resources necessary for extracting experimentally-relevant measurements. 

A recently introduced basis, the \textit{mixed basis}~\cite{DAndrea:2023qnr, Grabowska:2024emw}, serves as a middle ground between the irrep and group element representations of the gauge group. With the removal of gauge-redundant degrees of freedom, as well as the use of a coupling-dependent digitization scheme, this basis is believed to provide a resource-efficient formulation of \sutwo for all values of the gauge coupling. This is of particular importance when extracting physical parameters from numerical simulations, due to the necessary tunings of bare couplings to reach the appropriate large volume continuum limits. 

The present work develops algorithms for simulating the real-time dynamics of the minimal non-trivial example of a fully gauge-fixed \sutwo Hamiltonian: two plaquettes with open boundary conditions. Sec.~\ref{sec:theory_section} provides a short review of gauge theory Hamiltonians and gauge fixing. It also details the group element, mixed and character irrep bases, each of which is optimized for distinct coupling regimes.  Sec.~\ref{sec:digitization} analyzes the mixed basis in detail, presenting the corresponding digitization scheme and mapping of the Hamiltonian into quantum circuits for time evolution. It also provides resource estimates for larger-scale simulations. Sec.~\ref{sec:MixedBasisCircuits} gives two different methods for translating the digitization into quantum circuits that can be implemented onto Noisy Intermediate-Scale Quantum (NISQ)-era devices. Finally, Sec.~\ref{sec:quantum_results} presents the results of a real-time simulation on \textbf{ibm\_fez}, a 156 superconducting qubit chip. An in-depth explanation of the various error-mitigation strategies that are utilized in order to extract results from the noisy quantum simulation is also presented. Finally, Sec.~\ref{sec:Conclusions} discusses the implications of this work and proposes directions for future work and extensions to larger system sizes.
\begin{figure*}[t]
    \centering
    \includegraphics[scale=.5, trim = 0 -10 0 10]{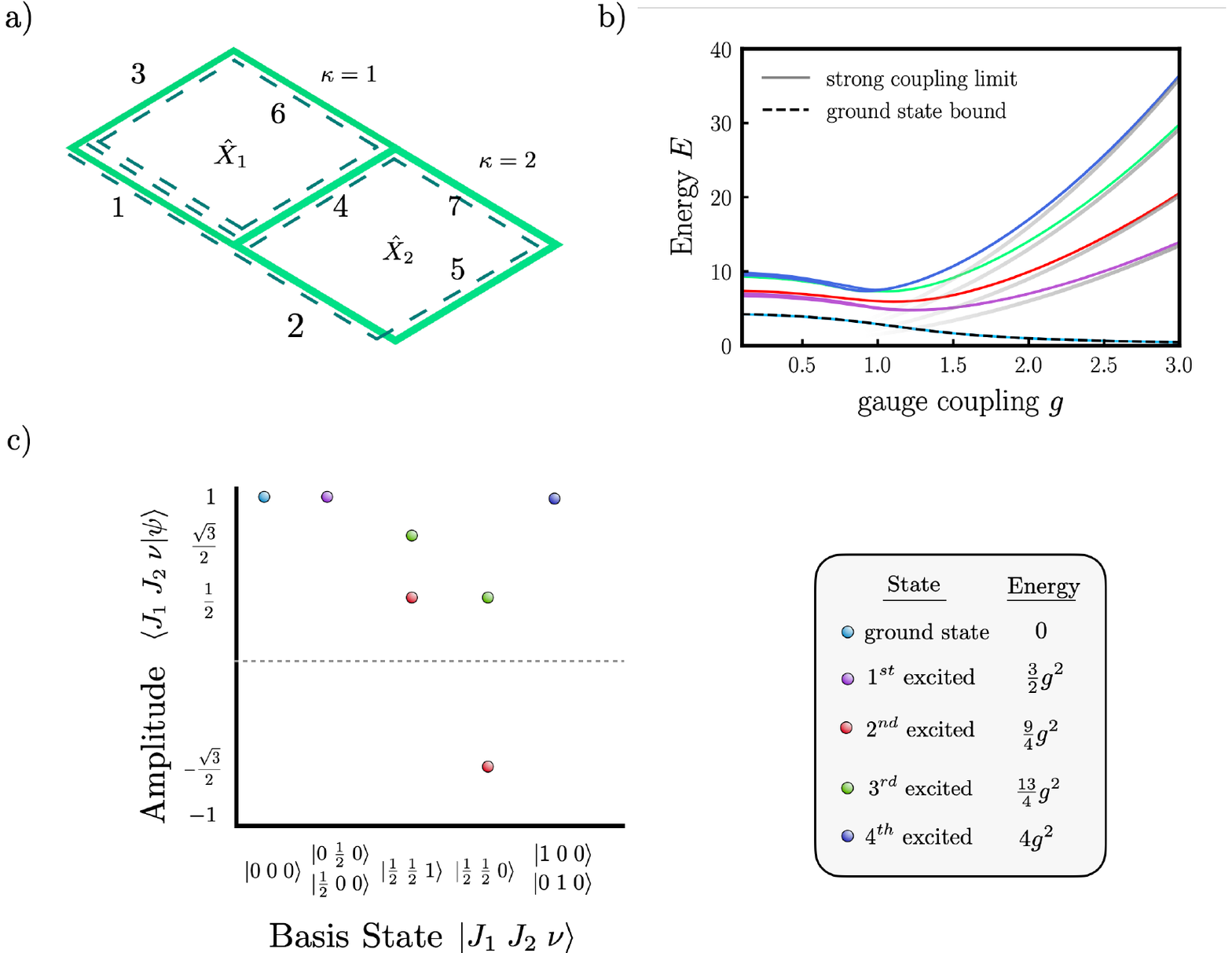}
    \caption{\textit{Spectrum of the Two Plaquette System} \textbf{a)} Two plaquette system with open boundary conditions. The system contains seven links, numbered $1$ through $7$, with the physical links denoted as $\kappa=1,2$. The magnetic degrees of freedom $\hat{X}_{1},\hat{X}_2$ are associated with each plaquette. The dashed lines that run around each plaquette denote the loops used in the max tree gauge fixing procedure. \textbf{b)} Spectrum of the Hamiltonian for different bare couplings. The solid lines show the solution obtained by FEM. The black dashed line on the bottom shows the energy bound of Eq.~\eqref{eq:bound}. The grey lines denote the strong coupling limit of the character irrep spectrum. This limit is found by ignoring the contribution of the magnetic Hamiltonian, which is only valid when the gauge coupling is large. \textbf{c)} Low-lying spectrum of the character irrep Hamiltonian in the strong coupling limit. The first and fourth excited state are two-fold degenerate and each state is a linear superposition of the stacked basis states. See Table~\ref{tab:CharIrrepStrong} for further details.}
    \label{fig:two_plaq_description}
\end{figure*}
\section{Two Plaquette Hamiltonian}\label{sec:theory_section}
This section provides a concise overview of how to formulate lattice gauge theories using maximal-tree gauge fixing~\cite{PhysRevD.15.1128}, as well as how to remove the remaining global gauge transformation. It also introduces several different bases that can be used to derive analytic results for a wide range of (bare) coupling values. For a pedagogical review about formulating gauge theories on the lattice and understanding gauge symmetries in that construction, see Ref.~\cite{DAndrea:2023qnr}.

\subsection{Gauge Fixing Lattice Gauge Theories}
The Kogut-Susskind Hamiltonian~\cite{PhysRevD.11.395} is the starting point of this work. Written in terms of electric field link operators and magnetic plaquette operators, the general form of a lattice gauge Hamiltonian is
\begin{align}
\hat H = \frac{g^2}{2a}\sum_{\ell \in \text{links}}\hat E^2_\ell+ \frac{1}{2g^2a}\sum_{p \in \text{plaq.}}\Tr\left[2 - \hat P^{\phantom{\dagger}}_p-\hat P^\dagger_p\right]
\end{align}
where each plaquette is constructed out of gauge link operators, $g$ is the gauge coupling, and $a$ is the lattice spacing, typically set to the identity. Before introducing the commutation relations of the operators that appear in this Hamiltonian, it is beneficial to understand how these operators are related to the mathematics of Lie groups. 

Typically, a gauge link operator is defined as the path-ordered Wilson line of the gauge field,
\begin{align}
U_{rs}(\ell) \equiv \left[\CP \exp \left(i \int_{\boldsymbol{n}}^{\boldsymbol{n}+\boldsymbol{e}_i}dx^\mu A_\mu(x)\right)\right]_{rs}
\end{align}
where the gauge link $\ell(\boldsymbol{n}, \boldsymbol{e}_i)$ originates at point $\boldsymbol{n}$ and extends in the direction $\boldsymbol{e}_i$ and $r,s$ are color indices. The Hilbert space of this operator is the space of all square-integrable wavefunctions $L^2(G, \mathfrak{g})$ over the Lie group $G$ with respect to the Haar measure $d\mathfrak{g}$. Therefore, any gauge link operator can be written in the group element basis as
\begin{align}
\hat U_{rs}(\ell) \equiv \int d\mathfrak{g}_\ell \, u_{rs}(\mathfrak{g}_\ell)\ket{\mathfrak{g_\ell}}\bra{\mathfrak{g_\ell}}
\end{align}
where $u_{rs}$ are the matrices that define the group elements in the fundamental representation. The full Hilbert space is simply the tensor product of the local dimensions of each link.

The electric operators are also defined in this language. In particular, the electric operators are responsible for translations in the space of group elements,
\begin{align}
\hat \Theta_{L\mathfrak{h}}= e^{i \phi^a(\mathfrak{h})\hat E^a_L}\qquad \hat \Theta_{R\mathfrak{h}}= e^{i \phi^a(\mathfrak{h})\hat E^a_R} 
\end{align}
where $\phi^a(\mathfrak{h})$ are parameters typically called normal coordinates, where $a\in\{1,\dots,\text{dim}(G)\}$. The operators $\hat\Theta$ act on the group element basis states as
\begin{align}
\hat \Theta_{L\mathfrak{h}}\ket{\mathfrak{g}}= \ket{\mathfrak{h}^{-1}\mathfrak{g}} \qquad \hat \Theta_{R\mathfrak{h}}\ket{\mathfrak{g}}= \ket{\mathfrak{g}\mathfrak{h}^{-1}} \, ;
\end{align}
since the left and right electric field operators act on states differently, both need to be present for non-Abelian gauge theories.

The electric field operators satisfy the commutation relations
\begin{gather}
\left[\hat E_L^a, \hat E_L^b\right]= - i f^{abc}\hat E_L^c \qquad \left[\hat E_R^a, \hat E_R^b\right]=  i f^{abc}\hat E_R^c \nonumber \\
\left[\hat E_L^a, \hat E_R^b\right] = 0 \, 
\end{gather}
and given that the left and right electric operators commute, they each furnish an independent Lie algebra. Additionally, each electric operator has a commutation relation with the gauge link operator
\begin{align}
\left[\hat E^a_{L}, \hat U^{(r)}_{mn}\right] = T^{(r)a}_{mm'}\hat U^{(r)}_{m'n} \quad \left[\hat E^a_{R}, \hat U^{(r)}_{mn}\right] = \hat U^{(r)}_{mn'}T^{(r)a}_{n'n}
\end{align}
where $r$ indicates the representation and is \textit{not} summed over. 

So far, there has been no mention of gauge invariance in this discussion, though all the ingredients are now in place. For a theory that is gauge invariant, there exist states that are equivalent to each other; these states are related via gauge transformations. In deriving the Kogut-Susskind Hamiltonian from the Wilson action, Weyl gauge is imposed. This is an incomplete gauge-fixing condition and purely spatial gauge transformations still remain. A gauge transformation $\Omega$ at the lattice site $\boldsymbol n$ is performed by applying the operator
\begin{align}
\hat \Theta_{\Omega}(\boldsymbol{n}) \equiv \exp\left(i \phi^a(\Omega)\hat G^a(\boldsymbol{n})\right)
\end{align}
where $\hat G^a$ is the Hermitian operator
\begin{align}
\hat G^a(\boldsymbol{n}) = \sum_{i=1}^d \left[\hat E^a_R(\boldsymbol{n}- \boldsymbol{e}_i, \boldsymbol{e}_i)-\hat E^a_L(\boldsymbol{n}, \boldsymbol{e}_i)\right]
\end{align}
and is commonly called the Gauss Law operator. When this operator acts on any state in the physical subspace,
\begin{align}
\hat G(\boldsymbol{n}) \ket{\Psi_\text{phys}} = 0 \quad \forall\, \boldsymbol{n} \, 
\end{align}
is satisfied. It is important that these equivalent states are not treated as distinct. One possible approach for avoiding this is gauge-fixing, a method for choosing a single representative state from each set of equivalent states.

Gauge fixing on a spatial lattice can be done systematically, without the need to solve the Gauss law equation at each site. Instead, the residual spatial gauge transformations that are still present after Weyl gauge-fixing are used to set certain gauge links to the identity. Every time a gauge transformation is used to fix a specific gauge link, neighboring links are also transformed. In two or more spatial dimensions, it is not possible to set all the gauge links to the identity, which is consistent with gauge bosons being dynamical in these dimensions. Ref.~\cite{PhysRevD.15.1128} showed that the largest set of gauge links that can be set to the identity are those that make up a maximal tree.\footnote{A maximal tree is a set of all gauge links such that if any additional link was added to the set, the lattice would now contain a closed loop. See Ref.~\cite{DAndrea:2023qnr} for more details} This convention removes all of the residual spatial gauge transformations except for those at a single lattice site, which is typically taken to be the origin. This last spatial gauge transformation, which is a global gauge transformation, must be treated differently. Once all the spatial gauge redundancy is fully removed, the resultant Hamiltonian only spans the physical subspace and there are no more equivalent states.

This process has been completed for \sutwo lattice gauge Hamiltonians in two- and three-dimensions in Ref.~\cite{DAndrea:2023qnr,Grabowska:2024emw}. In the first work, the Kogut-Susskind Hamiltonian is gauge-fixed using the maximal tree gauge fixing procedure and rewritten in terms of new canonically-conjugate operators, $\hat{\cal E}^a_{L,R}(\kappa)$ and $\hat X(\kappa)$. The operators $\hat{\cal E}^a_{L,R}(\kappa)$ are related to parallel-transported electric link operators and $\hat X(\kappa)$ are related to parallel-transported gauge links; both sets of operators are transported along a path $P(\kappa)$. In terms of these variables, the Hamiltonian is given by
\begin{align}
\label{eq:H_maxTree}
    \hat H_E &= \frac{g^2}{2 a} \sum_{\ell} \left( \sum_{\kappa \in t_+(\ell)} \hat{\mathcal{E}}^a_{L}(\kappa) - \sum_{\kappa \in t_{-}(\ell)} \hat{\mathcal{E}}^{a}_{R}(\kappa) \right)^2\nn\\
    \hat H_B &= \frac{1}{2g^2 a}\sum_p \Tr \left( \hat I - \prod_{\kappa\in p} \hat X(\kappa)^{\sigma(\kappa)} \right) + \mathrm{h.c.}
\end{align}
where $\sum_p$ and $\sum_\ell $ denotes sums over all plaquettes and all links of the lattice, $t_\pm(\ell)$ is the set of all physical links such that $\ell$ is contained in the path $P$ as a positive/negative link and $\sigma = \pm 1$, depending on whether the link is traversed in the positive or negative orientation. The precise form of the Hamiltonian depends on the chosen max-tree and path convention; physical results will be convention-independent.

In addition, Ref.~\cite{DAndrea:2023qnr} provides a group-element basis for constructing the Hilbert space of the Hamiltonian of Eq.~\eqref{eq:H_maxTree}. Using the axis-angle representation of rotations in three-dimensional space, each loop variable is represented by a two-dimensional matrix
\begin{align}
\label{eq:XOrig}
\hat X &=\left(
\begin{array}{cc}\cos \frac{\omega}{2}-i \cos \vartheta \sin \frac{\omega}{2} & -i e^{-i \varphi} \sin \vartheta \sin \frac{\omega}{2} \\
-i e^{i \varphi} \sin\vartheta \sin \frac{\omega}{2}  & \cos \frac{\omega}{2}+i \cos\vartheta \sin \frac{\omega}{2} \\
\end{array}\right) 
\end{align}
where the variables have ranges
\begin{align}
\omega \in [0, 2\pi],\;\;\theta \in [0, \pi],\;\;\phi \in [0, 2\pi] \,\, .
\end{align}
The corresponding canonically-conjugate variables are given by
\begin{align}
\hat{\boldsymbol{{\cal E}}}_{R/L} = \frac{1}{2}\left(\hat{\boldsymbol{\Sigma}} \pm \hat{\boldsymbol{L}}\right) \, .
\end{align}
In axis-angle coordinates, $\hat{\boldsymbol{L}}$ is given by
\begin{align}
\label{eq:Lop}
\hat L^x &= i \left( \sin\varphi \frac{\partial}{d \vartheta} + \cot\vartheta \cos\varphi \frac{\partial}{{\partial} \varphi}\right)\nonumber\\
\hat L^y &= i \left( -\cos\varphi \frac{\partial}{{\partial} \vartheta} + \cot\vartheta \sin\varphi \frac{\partial}{{\partial} \varphi}\right)\nonumber\\
\hat L^z &= -i \frac{\partial}{{\partial} \varphi} \, ,
\end{align}
which is simply the (conventional) angular momentum operator. The operator $\hat{\boldsymbol{\Sigma}}$ can be written in terms of $\hat{\boldsymbol{L}}$ and the position vector $\hat{\boldsymbol{\eta}}$ as
\begin{align}
\hat{\boldsymbol{\Sigma}} = 2 i \hat{\boldsymbol{\eta}} \pdv{}{\omega}+ \cot \frac{\omega}{2}\left(\hat{\boldsymbol{\eta}} \cross \hat{\boldsymbol{L}} \right) \, ,
\end{align}
where $\hat{\boldsymbol{\eta}}$ is the axis of rotation in three-dimensional space,
\begin{align}
\label{eq:EtaDef}
\hat{\boldsymbol{\eta}} \equiv \left(\sin \vartheta \cos \varphi,\sin \vartheta \sin \varphi, \cos \vartheta\right) \,\, .
\end{align}
The operator $\hat{\boldsymbol{\Sigma}}$ is analogous to the boost operator of the Lorentz group. The commutation relations of these two operators are 
\begin{align}
\left[\hat L^a, \hat L^b\right] &= i \epsilon^{abc}\hat L^c \nonumber \\
\left[\hat \Sigma^a, \hat \Sigma^b\right] &= i \epsilon^{abc}\hat L^c \nonumber \\
\left[\hat \Sigma^a, \hat L^b\right] &= i \epsilon^{abc}\hat \Sigma^c \, ,
\end{align}
whereas the commutation relations for the Lorentz group operators have a relative minus sign for the second and third commutation relation.
The Haar measure for each link in this basis is
\begin{align}
d\mathfrak{g}= 4 \sin^2\frac{\omega}{2}\sin \vartheta \, d \omega\, d\vartheta\, d\varphi
\end{align}

The fixing of the last (global) gauge transformation is carried out in Ref.~\cite{Grabowska:2024emw}. The key insight of that work is that the Hamiltonian can be conceptualized as a system of rods all connected together at the origin. The $\kappa^\text{th}$ rod has length $\omega_\kappa$ and its orientation is given by the polar angle $\vartheta_\kappa$ and azimuthal angle $\varphi_\kappa$. Such a system can be written either in a lab-frame coordinate system or in a body-frame coordinate system. An Euler rotation relates these two coordinate systems. In the lab-frame coordinate system, the location of each rod is parameterized as in Eq.~\eqref{eq:EtaDef}. This same configuration of rods can be written using body-frame variables, coupled with an Euler rotation. The definition of the body frame is arbitrary, but Ref.~\cite{Grabowska:2024emw} chooses to define the $\hat z$ axis of the body frame by aligning it with one rod and the $\hat x -\hat y$ plane via the location of another rod. In this coordinate system, the location of each rod is given by
\begin{align}
\boldsymbol{\eta}_1 &= \boldsymbol{R}\cdot \left(0, 0, 1\right) \\
\boldsymbol{\eta}_2 &= \boldsymbol{R}\cdot \left(\sin\Theta, 0, \cos \Theta\right) \nonumber \\
\boldsymbol{\eta}_\mu &= \boldsymbol{R}\cdot\left(\sin \theta_\mu \cos \phi_\mu,\sin \theta_\mu \sin \phi_\mu, \cos \theta_\mu\right) \qquad \mu \geq 3  \nonumber
\end{align}
where $\boldsymbol{R}$ is the rotation matrix written in terms of Euler angles. Rotation of the full system leaves the lengths of each rod invariant. In carrying out the change of basis to go from the lab-frame coordinate system to the body-frame coordinate system, the remaining global gauge transformation is made visible. The Hamiltonian can then be trivially fixed to the total gauge-singlet sector and only depends on the variables $\{\omega_\kappa, \Theta, \theta_\mu, \phi_\mu\}$; all dependence on the Euler angles is eliminated from the Hamiltonian. Ref.~\cite{Grabowska:2024emw} provides the explicit forms of all possible electric bilinears as well as magnetic loop operators in this new coordinate system. The new basis is called the `sequestered basis', as all the dependence on the total charge of the system is \textit{sequestered} into the Euler angles. The focus of this paper is to simulate the two plaquette version of this system, on a NISQ-era device.

\subsection{Two Plaquette Hamiltonian}
Using Eq.~\eqref{eq:H_maxTree} and a loop convention where each loop only encompasses one plaquette, the Hamiltonian for the two plaquette system with open boundary conditions, shown in Fig.~\SubFigRef{fig:two_plaq_description}{a}, is given by
\begin{align}
\label{eq:HTPOp}
\hathtp &=\frac{g^2}{2}\left(4\, \hat{{\cal E}}_1^2+6\, \hat{{\cal E}}_2^2 - 2\, \hat{\boldsymbol{{\cal E}}}_{R2}\cdot \hat{\boldsymbol{{\cal E}}}_{L2} \right.\nonumber \\
&\left.+ 2\,\hat{\boldsymbol{{\cal E}}}_{R1}\cdot \hat{\boldsymbol{{\cal E}}}_{R2} -4\,\hat{\boldsymbol{{\cal E}}}_{R1}\cdot \hat{\boldsymbol{{\cal E}}}_{L2}\right)\nonumber \\
&+\frac{1}{g^2}\left(4-\tr \hat X_1-\tr \hat X_2 \right) \, .
\end{align}
The lattice spacing has been set to the identity, as is conventionally done, and
\begin{align}
\hat{\boldsymbol{{\cal E}}}_{L \kappa} \cdot \hat{\boldsymbol{{\cal E}}}_{L \kappa}=\hat{\boldsymbol{{\cal E}}}_{R \kappa} \cdot \hat{\boldsymbol{{\cal E}}}_{R \kappa}\equiv  \hat{{\cal E}}_\kappa^2 \, ,
\end{align}
where $\kappa$ is not summed over.
A basis that spans the Hilbert space of $\hathtp$ must be defined to analyze the real-time behavior of this system. Additionally, both $\hat{\cal E}_\kappa$ and $\hat X$ are infinite-dimensional operators and the basis needs to be truncated. The calculation of any physical observable is basis-independent but truncation will affect this universality. The effects should disappear as the truncation is removed. This paper utilizes three different bases to measure both static and dynamic properties.

\subsubsection{Group Element Basis}\label{subsub_sec:groupel_basis}
In the group element basis, the eigenstates $\ket{\mathfrak{g}}$ of the gauge link operators (or the loop operators) are the basis states. Any parameterization of $\ket{\mathfrak{g}}$ can be used for defining the action of the Hamiltonian on the Hilbert space but some parameterizations are more convenient than others. For \sutwo, two natural parameterizations are Euler angles or axis-angle coordinates. The axis-angle parameterization is particularly convenient~\cite{DAndrea:2023qnr,Grabowska:2024emw}. Any wavefunction that spans only the physical Hilbert space \textit{i.e.} any gauge-singlet wavefunction is given by
\begin{align}
\label{eq:RescalePsi}
\braket{\omega_1\, \omega_2\, \Theta}{\Psi} &= \Psi(\omega_1,\omega_2, \Theta)\nonumber \\
&\equiv \frac{u(\omega_1, \omega_2, \Theta)}{4\sin\frac{\omega_1}{2}\sin\frac{\omega_2}{2}}
\end{align}
where the rescaled wavefunction $u(\omega_1, \omega_2, \Theta)$ is introduced to simplify later calculations. The magnetic component of the fully-gauged fixed  Hamiltonian is diagonal and only depends on the radial coordinates $\omega_{1}$ and $\omega_2$,
\begin{align}
\label{eq:HBGE}
 H^{(B)}_{\square\square}= \frac{2}{g^2}\left(2- \cos \frac{\omega_1}{2}-\cos \frac{\omega_2}{2}\right) \,.
\end{align}
The electric component of the Hamiltonian can be determined by evaluating each of the electric bilinears that appear in Eq.~\eqref{eq:HTPOp}. The explicit expressions for the electric bilinears are relegated to Appenix~\ref{app:GEB}. Combining the various elements together, the electric component of the Hamiltonian is given by
\begin{align}
\label{eq:HEGE}
\htp^{(E)}&=-\frac{g^2}{2}\left[\left\{4\left(\pdv[2]{}{\omega_1}+ \cot \frac{\omega_1}{2}\pdv{}{\omega_1}\right)-2 \csc^2\frac{\omega_1}{2}\CN\right.\right.\nonumber \\
&+\left.\left( \cot \frac{\omega_2}{2}\pdv{}{\omega_1}\right)\left(\sin \Theta\pdv{}{\Theta}\right)+\left(\omega_1 \leftrightarrow \omega_2\right)\right\}  \nonumber \\
&+ \left(\frac{1}{2}\cot \frac{\omega_1}{2}\cot \frac{\omega_2}{2}\right)\left(\sin \Theta\pdv{}{\Theta}\right)\nonumber \\
&-\left.\left(\frac{1}{2}\cos \Theta \cot\frac{\omega_1}{2}\cot\frac{\omega_2}{2}-\frac{1}{2}\right)\CN\right]
\end{align}
where $\CN$ is the differential operator,
\begin{align}
\label{eq:LegDifEq}
\CN = - \pdv[2]{}{\theta}- \cot \theta \pdv{}{\theta} \, .
\end{align}

In this basis, $\htp$ is a partial differential operator and its spectrum can be found numerically using Finite Element Methods (FEM). FEM is a general numerical approach for solving low-dimensional partial differential equations that utilizes a discrete mesh of the computational domain.\footnote{This is in contrast to Finite Difference Methods (FDM), which approximate differential operators as finite differences on discrete grid.} In order to ensure that the eigenfunctions of $\htp$ are normalizable and that the mesh does not encounter any singularities, FEM is applied to solve for the rescaled wavefunctions, $u(\omega_1, \omega_2, \Theta)$. In this case, the radial coordinates, $\omega_{1,2}$ have a range of $[0, 2\pi]$ with Dirichlet boundary conditions while the angular coordinate $\Theta$ has range $[0, \pi]$ with Neumann boundary conditions. Fig.~\SubFigRef{fig:two_plaq_description}{b} plots the low-energy spectrum as a function of the bare gauge coupling. These solutions provide a beneficial `exact' solution to which the mixed basis and character irrep basis can be compared. This is particularly helpful, as while the these bases can be used for larger system sizes, the FEM approach faces insurmountable computational barriers when applied to larger systems. Without fully gauge-fixing, FEM could not even be applied to the two-plaquette system as, for example, the (naively parameterized) two-plaquette version of the Kogut-Susskind Hamiltonian involves twenty-one parameters (three quantum numbers for each of the seven links). 

\subsubsection{Mixed Basis}\label{subsub_sec:mixed_basis}
In the mixed basis, the states that span the Hilbert space are labeled by the continuous variables $\omega_1, \omega_2$ and the discrete quantum number $\nu$. This discrete quantum number is related to the angular variable $\Theta$ via
\begin{align}
\braket{\Theta}{\nu} = P_\nu(\Theta)
\end{align}
with $P_\nu(\Theta)$ the (rescaled) Legendre polynomial of degree $\nu$.\footnote{The conventional normalization of the Legendre polynomials is 
\begin{align}
\int_0^\theta d(\cos \theta)\,P_{\nu'}(\theta)P_\nu(\theta) = \frac{2}{2\nu+1}\delta_{\nu'\nu}
\end{align}
but due to the requirement that the basis states of the Hilbert space be orthonormal, the Legedre polynomials in this work are rescaled.} Legendre polynomials are used to span the Hilbert space because they are the eigenfunctions of the operator $\CN$,  
\begin{align}
    \CN P_\nu(\theta) = \nu(\nu+1) P_\nu(\theta) \, .
\end{align}

The matrix elements in this basis are found by calculating
\begin{align}
&\mel{\omega_1'\omega_2'\nu'}{\hathtp}{\omega_1\omega_2 \nu} = \int d(\cos \Theta) d(\cos\Theta')  \braket{\nu'}{\Theta'}\nonumber \\
&\hspace{50pt}\times \mel{\omega_1'\omega_2'\nu'}{\hathtp}{\omega_1 \omega_2 n}\braket{\Theta}{\nu} \nonumber\\
&\hspace{50pt}= \int d(\cos \Theta)  P^*_{\nu'}(\Theta)\htp P_\nu(\Theta)
\end{align}
where $\htp$ is the sum of the magnetic and electric Hamiltonians, written in the group element basis in Eq.~\eqref{eq:HBGE} and Eq.~\eqref{eq:HEGE}

To convert between the group element basis and the mixed basis, only four integrals need to be calculated:
\begin{align}
\int \,d(\cos\Theta)\, P_{\nu'}(\Theta)
\left\{\begin{array}{c}
1\\
\CN \\
\cos \Theta\\
\sin \Theta \pdv{}{\Theta}
\end{array}\right\}
P_{\nu}(\Theta) \, .
\end{align}
The results of these integrals, tabulated in Appendix~\ref{app:MB}, are used to recast expressions in $\htp$. The necessary digitization, particularly of the components that depend on $\omega_\kappa$, will be discussed below in Sec.~\ref{sec:digitization}.

Using the mixed basis, it is possible to derive what turns out to be a very strong bound on the ground state energy of the two plaquette system. Recall that the ground state energy of a Hamiltonian, $E_0$, can be bounded from above by the expectation value of that Hamiltonian in any state $\Psi$,
\begin{align}
E_\Psi \equiv \mel{\Psi}{\hat H}{\Psi} \geq E_0 \, .
\end{align}
In general, the goodness of this bound depends on how well the wavefunction $\Psi$ matches the true ground state. A good ansatz for the two plaquette system is found by assuming that the ground state does not depend strongly on $\Theta$ and therefore the functional dependence of the ground state on $\omega_\kappa$ is encapsulated in the differential operator
\begin{align}
-2g^2\left(\pdv[2]{}{\omega_\kappa}+ \cot \frac{\omega_\kappa}{2}\pdv{}{\omega_\kappa}\right)+\frac{2}{g^2}\left(1-\cos \frac{\omega_\kappa}{2}\right) \, ,
\end{align}
which is the component of the Hamiltonian that is diagonal in $\nu$ for each $\omega$. This differential equation is related to that of the Mathieu equation. Therefore, the ansatz for the the ground state is
\begin{align}
\label{eq:GSAnsatz}
u(\omega_1, \omega_2, \Theta) = \frac{\text{se}_2(-8/g^4, \omega_1/4)\text{se}_2(-8/g^4, \omega_2/4)}{\sqrt{2}} \,
\end{align}
where $\Psi$ and $u$ are related via Eq.~\eqref{eq:RescalePsi}. The parity odd Mathieu function, $\text{se}_n(q, x)$ has period  $\pi$ when $n$ is even or $2\pi$ when $n$ is odd. For this system, $n$ has to be even for $\psi$ to be finite on the full range of $\omega_\kappa$.  With this ansatz, $E_\Psi$ is 
\begin{align}\label{eq:bound}
E_\Psi &= \frac{g^4 b_2\left(-8/g^4\right)-4 g^4+16}{4 g^2} \nonumber \\
&=\begin{cases}
3\sqrt{2}&\qquad g\ll 1 \\
4/g^2 &\qquad g \gtrsim 1
\end{cases}
\end{align}
where $b_n(q)$ is the characteristic number that guarantees that $\text{se}_n(q, x)$ has period $\pi$ or $2\pi$. Comparing this ansatz to the FEM solution demonstrates that there is very good agreement between the two for a wide range of gauge couplings.

\subsubsection{Character Irrep Basis}
\label{subsubsec:CharIrrep}
The Kogut-Susskind Hamiltonian is formulated in the group irrep basis, also called the electric basis. The eigenstates of the single link Hilbert space are $\ket{J M_L M_R}$ where $J$ is the total angular momentum and $M_L$ ($M_R$) is the angular momentum projected onto the $\hat z$ axis of the lab (body) frame; $J$ is also commonly referred to as the irrep of \sutwo group. In analogy, the basis of the Hilbert space of the fully-gauged fixed two-plaquette Hamiltonian can be chosen to be $\ket{J_1\, J_2\, \nu}$ where $J_{1,2}$ are again irreps of \sutwo. The relationship between the group element basis and this basis is
\begin{align}
\braket{\omega_1\,\omega_2\,\Theta}{J_1\,J_2\,\nu} = \chi^{J_1}_\nu(\omega_1)\chi^{J_2}_\nu(\omega_2)P_\nu(\Theta)
\end{align}
where $\chi^J_\nu$ are (rescaled) generalized character functions of the irreps of the rotation group. These functions satisfy the differential equation,
\begin{align}
\left[\pdv[2]{}{\omega}+\cot \frac{\omega}{2}+J(J+1)-\frac{\nu(\nu+1)}{4 \sin^2\frac{\omega}{2}}\right]\chi^J_\nu(\omega) = 0
\end{align}
where $J$ are integers and half-integers and $\nu$ are integers, bounded by $2J\geq \nu\geq 0$. An explicit construction of these character functions is done via Clebsch-Gordan coefficients,
\begin{align}
\chi_\nu^J(\omega) = \frac{i^\lambda}{2\sqrt{\pi}} \sum_{M}e^{-i M \omega}C^{JM}_{JM\nu 0}
\end{align}
where the prefactor of $1/(2\sqrt{\pi})$ arises due to the requirement that the basis states are orthonormal.

Relevant matrix elements can be calculated for generic $J_1, J_2$ and $\nu$ by making use of recurrence relations for the character functions~\cite{khersonskii1988quantum}. The complete expressions are presented in Appendix~\ref{app:CIB}. A key result is that the magnetic component of the Hamiltonian only changes the $J_\kappa$ quantum numbers by at most $\pm 1/2$. In contrast, the electric component of the Hamiltonian only changes the $\nu$ quantum number by at most $\pm 1$. With these matrix elements, it is possible to construct a finite-dimensional Hamiltonian, with cutoff $J_{\kappa}^\text{max}$. Note that $\nu$ is naturally truncated due to the requirement that 
\begin{align}
2\,\text{Min}(J_1, J_2)\geq \nu \geq 0
\end{align}
and since the Hamiltonian is symmetric under $1 \leftrightarrow 2$ exchange, as is evident from the differential form of the Hamiltonian, both $J_1$ and $J_2$ have the same cut-off. 

Solving for the spectrum in the strong coupling limit is simple, as the electric component preserves the quantum numbers $J_1, J_2$ and the contributions from the magnetic component can be ignored. Additionally, since the quantum number $\nu$ is bounded from above by $2\,\text{Min}(J_1,J_2)$, for a given $J_1, J_2$, the relevant sub-block of the Hamiltonian is finite-dimensional. In short, in the strong coupling limit, at leading order, the Hamiltonian is block-diagonal. The results for the lowest states are show in Fig.~\SubFigRef{fig:two_plaq_description}{c} and in Table~\ref{tab:CharIrrepStrong}.

This basis can be used for other values of the gauge coupling if the magnetic component of the Hamiltonian is included. In this case, the Hamiltonian is no longer diagonal in $J_{1,2}$ and its spectrum cannot be found analytically. Using exact diagonalization for different values of $J^\text{max}_\kappa$, it can be shown that at small values of the gauge coupling, the energy eigenstates diverge as $1/g^2$. However, using FEM, it is shown that the low-lying eigenstates become independent of $g$ in the limit of small $g$. These large truncation errors at weak coupling is why electric bases are disfavored for this regime.

\section{Digitization of the Mixed Basis}
\label{sec:digitization}
The Hilbert space degrees of freedom that may be represented by a register of qubits is finite and so a finite representation of the continuous gauge field must be determined. This is commonly referred to as a `digitization' (distinct from `discretization', which is corresponds to representing a continuous space-time on a lattice). This section focuses on the manner in which the mixed basis can be digitized, in particular the derivatives with respect to $\omega$. Of key importance is how to efficiently implement derivative operators onto a quantum device, which will be the focus of the next section.
\subsection{Conventions}
In the following sections, various ideas and tools will appear in slightly different contexts. Therefore, to ease notation and presentation, the following conventions are made.

For a single qubit, one may define the set of Pauli operators 
\begin{align}
\sigma_{\alpha}=\{I,X,Y,Z\} \,\,.
\end{align}
For $N$-qubits, the set of all Pauli strings (modulo phase) is the Pauli group, and is defined as \begin{align}
\mathcal{P}_N=\{\otimes_{l=1}^N\sigma_{l}^{\alpha_l}\;\; | \;\; \alpha_{l}=0,1,2,3\} \,\,.
\end{align}
This group defines a complete basis for the set of all hermitian operators $O$ (e.g. a Hamiltonian or density matrix) on $N$ qubits, and is normalized as 
\begin{align}
TrP P'=2^N\,\delta_{P,P'} \quad \forall\, P,P'\in\mathcal{P} \,\, .
\end{align}
Given such a hermitian operator $O$ on $N$ qubits, its (normalized) Pauli decomposition is denoted as 
\begin{align}
\mathcal{D}(O)=\left\{(P,\frac{1}{2^N}TrP O)\;\; | \;\; P\in \mathcal{P}_N\right\} \,\, .
\end{align}
$\mathcal{D}(O)$ will be used to either denote the set of pairs as shown above or the sum itself, where the meaning will be clear based on context.

\subsection{Formulating Gauge Field Operators}
An important feature of a good digitization is how quickly the observables (e.g. expectation values of hermitian operators) converge to the continuum limit as the resolution is increased. For fast convergence, the range of support of the wavefunctions of interest must be fully covered in both field and conjugate momentum space. In particular, if the full support of the wave function is not completely covered then predictions for observables will not converge to the correct solution. If the region of field space covered is much larger than the support of the wave function, then the digitization will sample regions of the wave function that are exponentially small and the convergence will be slow. The following digitization achieves good convergence properties by optimally spanning the regions of wave function support.

The continuous gauge field degrees of freedom represented by $\{\omega_1,\omega_2\}\in[0,2\pi)^{\otimes2}$ must be embedded into a register of $2n_q$ qubits with Hilbert space dimension $2^{2n_q}$. This is done by letting each $\omega_i$ take values on a grid of size $2^{n_q}$ with the range $[0,\omega_{\text{max}})$. The discretized variables are denoted as $\omega_i,$ for $i=1,2$ and assigned values 
\begin{align}
\omega_i &= \delta\omega(n+1/2) \qquad & n\in\{0,\dots,n_q-1\} \nonumber \\ \delta\omega &= \omega_{\text{max}}/n_q \, \, .
\end{align}
When performing classical simulations when the system is not embedded into qubits, the total grid size is denoted as $N_{\omega}$. In addition to the two $\omega$-registers, there is an additional quantum number $\nu$ that takes integer values and requires an additional $n_{\nu}$ qubits, giving a total simulation Hilbert space of size $2^{2n_q+n_{\nu}}$. This layout is graphically illustrated in Fig.~\SubFigRef{fig:digitization_details}{a}.

There are three types of terms that appear in the Hamiltonian, namely the first-order derivatives of $\omega_i$, the second-order derivates of $\omega_i$ and functions of $\omega_i$. Their action on the qubit register are treated separately. The field variable $\omega_i$ is defined by the operator
\begin{equation}
    \hat{\omega} = \frac{\delta\omega}{2}\left[n_qI-\sum_{j=0}^{n_q}2^j \hat Z_j\right]
\end{equation}
where $\hat Z_j$ is the Pauli $\hat Z$ operator acting on the $j$th qubit. Functions of $\hat{\omega}$ can then be constructed as $\mathcal{D}(f(\hat\omega))$, explicitly
\begin{align}
    f(\hat{\omega}) &= \sum_{P\in\{I,Z\}^{\otimes n_q}}f_{P} P\nonumber\\ 
    f_{P} &= \frac{1}{2^{n_q}}\text{Tr}[f(\hat{\omega})P]
\end{align}
where the sum above is over all strings of Pauli $Z$ operators. The second-order derivatives can be implemented using the exact lattice Laplacian with Dirichlet boundary conditions, which is given by
\begin{align}\label{eq:exact_laplacian}
\pdv[2]{}{\omega} &\mapsto \text{DST}^{-1}_{II}(-\hat{k}^2)\text{DST}_{II}\,\,
\end{align}
where $\text{DST}_{II}$ is a variation of the Discrete Sine Transform (DST), reflecting the boundary conditions in field space:
\begin{align}
    (\text{DST}_{II})_{kn} = 
    \begin{cases}
    \sqrt{\frac{2}{N}}\sin\frac{\pi(n+\frac{1}{2})(k+1)}{N}, & \text{for } k \neq N-1\\
    \sqrt{\frac{1}{N}}\sin\pi(n+\frac{1}{2}),  & \text{for } k = N-1
    \end{cases} \,\,.
\end{align}
Note that $\hat{k}$ is the lattice momentum operator, which is diagonal in Fourier space with eigenvalues $\pi k/\omega_{\text{max}}$ with $k\in\{1,\dots,N\}$. It has been shown that when using this exact formulation, the matrix elements of this term converge faster than polynomial to the continuum value as the grid spacing $\delta\omega$ goes to zero~\cite{Jordan:2011ci,Jordan:2017lea,Klco:2018zqz,Macridin:2018gdw}. Therefore it is generally advantageous to use compared to a finite difference.

The first-order derivative is, however, formulated using the finite central difference
\begin{align}\label{eq:finite_central_diff}
    \left(\pdv{}{\omega}\right)_{kn} = 
    \begin{cases}
    \frac{1}{2\delta\omega}, & \text{for } k-n=1,\\
    -\frac{1}{2\delta\omega}, & \text{for } k-n=-1,\\
    0 & \text{else}
    \end{cases}
\end{align}
The consequence of using the finite difference is that the digitized form of $\htp$ will only have polynomial convergence towards the continuum Hamiltonian as the spacing in field space is decreased. However, in the context of quantum simulation, even for  large $\delta\omega$ this error will generally be subleading to algorithmic errors stemming from approximations to continuous time evolution. While it is desirable to use spectral methods for the first-order derivative, as was done for the Laplacian, the Dirichlet boundary conditions complicate such an approach. This can be understood intuitively because the DST acts on data with odd symmetry, e.g. a sine wave. A first derivative  maps functions with odd symmetry to functions with even symmetry, and thus is incompatible with the use of a DST. Other types of spectral methods for first order derivatives with Dirichlet boundary conditions have been formulated, but often require non-uniform grid spacings, for example using Chebyshev nodes~\cite{Trefethen2013Approx}. This approach is undesirable as such grid choices have denser coverage towards the boundaries of the field region. This causes both oversampling of the singularities of the Hamiltonian as well as asymmetric sampling of the wavefunctions for low $g$.

\subsection{Optimal Gauge Field Truncation}

\begin{figure*}[t]
    \centering
    \includegraphics[width=\linewidth]{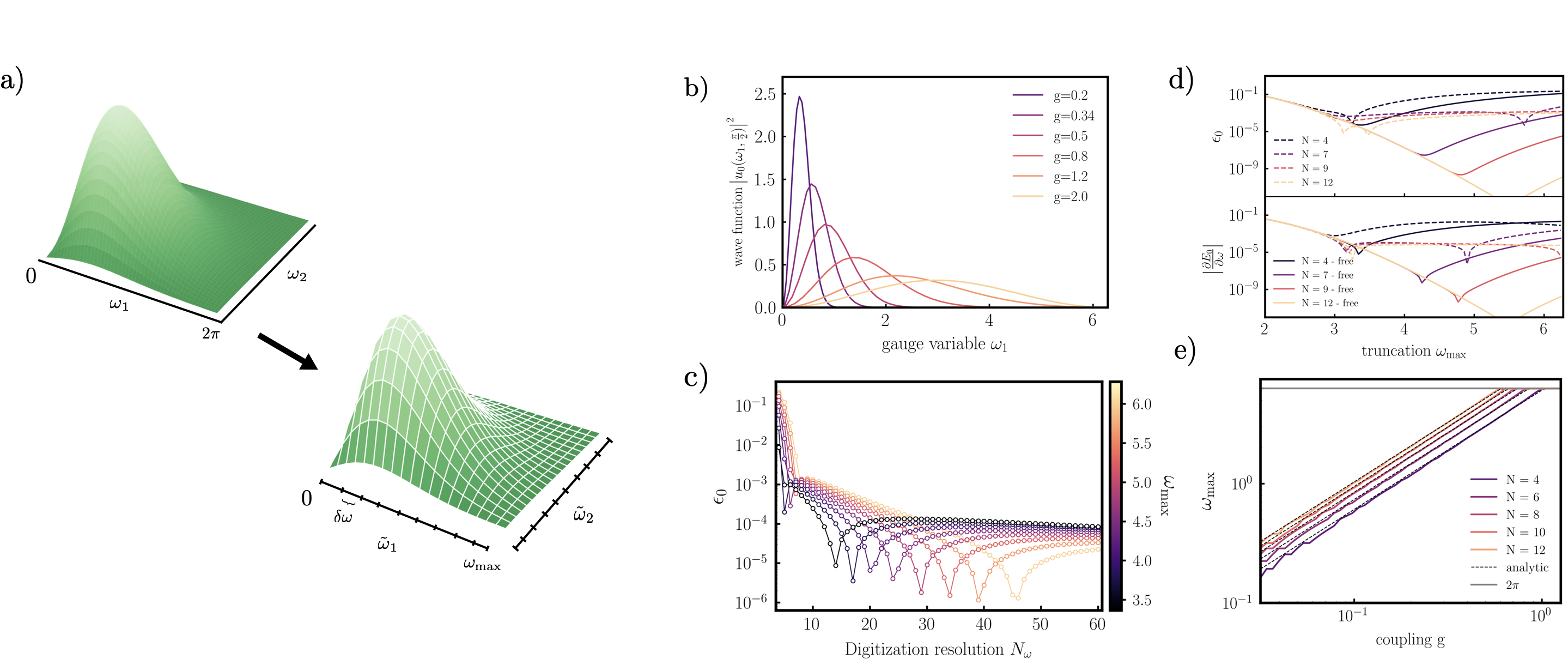}
    \caption{\textit{Digitization of Mixed Basis} \textbf{a)} The continuous range of $\omega$ is digitized on a discrete grid. As such the wave function will only be sampled at a discrete set of points. For the closest approximation to the undigitized theory, the maximum value of $\omega_{\text{max}}\leq 2\pi$ must be chosen. \textbf{b)} A slice of the ground state wave function for different values of the gauge coupling. As the gauge coupling decreases, the wave function becomes more concentrated around $\omega_1 = 0$, although it must go to $0$ at both boundaries due to the Dirichlet boundary conditions. \textbf{c)} Convergence of the ground state energy to the undigitized value as the digitization resolution $N_{\omega}$ is increased. The convergence is plotted for a variety of $\omega_{\text{max}}$. After quick convergence to $\epsilon_0\sim 10^{-3}$ a slower exponential approach occurs, followed by a plateau. \textbf{d)} \textit{(top)} Convergence to the ground state energy as $\omega_{\text{max}}$ is varied. The dashed lines correspond to the full two plaquette system while the solid lines correspond to the system without any first-derivative interactions. The effect of the first derivative reduces the precision that can be achieved. Even though this term limits precision, digitization errors are still below those due to time-evolution algorithms. \textit{(bottom)} The absolute value of the derivative of the energy w.r.t the truncation. It was heuristically found that the truncation at which this derivative takes a minimum corresponds to the optimal truncation for the analytically solvable case. This method is still applicable even when the exact solution is not known. \textbf{e)} The solution of the stationary condition for a range of gauge couplings. The agreement is very good with the analytical solution Eq.~\eqref{eq:OmegaMax}.}
    \label{fig:digitization_details}
\end{figure*}

In analogy with digitizations of scalar fields, the gauge field is truncated at some value $\omega_{\text{max}}\leq 2\pi$. Fig.~\SubFigRef{fig:digitization_details}{b} shows the amplitude of the ground state wave function
\begin{align}
u_0(\omega_1,\omega_2) = \left|\braket{\omega_1,\omega_2}{\psi_{\text{gs}}}\right|^2
\end{align}
along the $\omega_1$ axis ($\omega_2=\pi/2$) for different coupling strengths. As the gauge coupling is decreased, $u_0$ becomes increasingly localized at small $\omega_1$, decaying exponentially for large values. Therefore, the optimal $\omega_{\text{max}}$ must be characterized as a function of gauge coupling. This is done by considering the precision 
\begin{equation}\label{eq:precision}
    \epsilon_0=\left|1-\frac{E'_{0}}{E_0}\right|
\end{equation}
where $E'_0$ is the ground state energy predicted by the digitization and $E_0$ is the `true' ground state energy. Similar results are observed for low-lying eigenstates. The solutions for the digitization are compared to those found via FEM methods in Fig.~\SubFigRef{fig:two_plaq_description}{b}, the strong coupling solutions of Sec.~\ref{subsubsec:CharIrrep} and the ansatz of Eq.~\eqref{eq:GSAnsatz}, and are in good agreement. The solutions found via FEM are of particular use, and they have a relative precision of approximately $10^{-4}$.

Fig.~\SubFigRef{fig:digitization_details}{c} shows the precision for different choices of $\omega_{\text{max}}$ truncation as a function of digitization resolution $N_\omega$. The convergence is characterized by three stages. The first stage is rapid convergence to $\epsilon_0\sim 10^{-3}$ level for small $N_{\omega}$. Following this, the precision decays more slowly, though still exponentially, as $N_{\omega}$ is increased. Eventually, a plateau is reached. The precision value of this plateau decreases as $\omega_{\text{max}}$ increases. There are sharp spikes observed before the plateau is reached, which are attributed to the digitized solution ``accidentally" crossing the FEM solution, as opposed to approaching it. Such crossings are purely incidental and should not be interpreted as signs of optimality in representation of the wavefunction. Since the precision of the FEM solution is limited, it is not possible to study convergence below precision levels of $10^{-4}$. Despite this it is clear that even for grid sizes as low as $N_{\omega}=8\;(n_q=3)$ it is possible to reach $\epsilon_0\sim 10^{-3}$ levels of precision. This is at least an order of magnitude below algorithmic error due to time-evolution and on par with device noise-levels for near-term devices.

The optimal truncation is found by scanning over a range of $\omega_{\text{max}}$ values for a fixed $g,N_{\omega}$ and calculating the corresponding $\epsilon_0$. The top plot of Fig.~\SubFigRef{fig:digitization_details}{d} shows the precision $\epsilon_0$ for $\htp$. Also considered is the `free' effective theory of $\htp$, where terms containing a first derivative are subtracted off. This theory is studied to understand how $\epsilon_0$ behaves without the error induced by first derivatives. The ground state of the latter Hamiltonian can be found exactly and is given by the ansatz in Eq.~\eqref{eq:GSAnsatz}. The corresponding energy is Eq.~\eqref{eq:bound}. For the system without first derivatives, the precision decays exponentially in $\omega_{\text{max}}$ until the optimal value is reached, and then increases due to over-sampling regions of field space where the wave function has no support. However, once first derivatives are included, the precision plateaus at a value orders of magnitude higher, and can only be increased by increasing $N_{\omega}$. Thus there is a range of choices of $\omega_{\text{max}}$. The sharp dips are again due to accidental crossings and should be ignored when considering the optimality of $\omega_{\text{max}}$.

The gauge field truncation can be optimized without knowing the exact value of the ground state energy by looking at the derivative of the energy with respect to $\omega_{\text{max}}$. It is empirically observed that the optimal $\omega_{\text{max}}$ can equivalently be defined by the stationary condition
\begin{equation}
\label{eq:stationary_condition}
    \omega_{\text{max}} = \argmin_q\left|\frac{\partial E_i(\omega)}{\partial \omega}\right|\,\,.
\end{equation}
This condition can also be observed for the digitization of the scalar field. Fig.~\SubFigRef{fig:digitization_details}{d} (bottom) shows the derivative of the ground state energy predicted by the digitization. For the system without first derivatives, the minimum of these curves align with the points where the ground state energy most closely approaches the expression Eq.~\eqref{eq:bound}. When first derivatives are included, the precision plateaus at low values for a range of $\omega_{\text{max}}$. The optimal truncation in the absence of first derivatives is found by scanning over a range of gauge couplings and finding the solution of stationary condition, Eq.~\eqref{eq:stationary_condition}. Fig.~\SubFigRef{fig:digitization_details}{e} shows that the numerical solution found by the scan closely matches the analytical solution~\cite{DAndrea:2023qnr}
\begin{equation}
    \omega_{\text{max}} = \text{min}\left(g(N-1)\sqrt{\frac{\sqrt{8}\pi}{N}},2\pi\right)
    \label{eq:OmegaMax}
\end{equation}
which was found by minimizing $\langle[\partial/\partial\omega,\omega]-1\rangle$ for the ground state of a single plaquette.

\begin{figure}
    \centering
    \includegraphics[scale=.35, trim = 100 0 0 0]{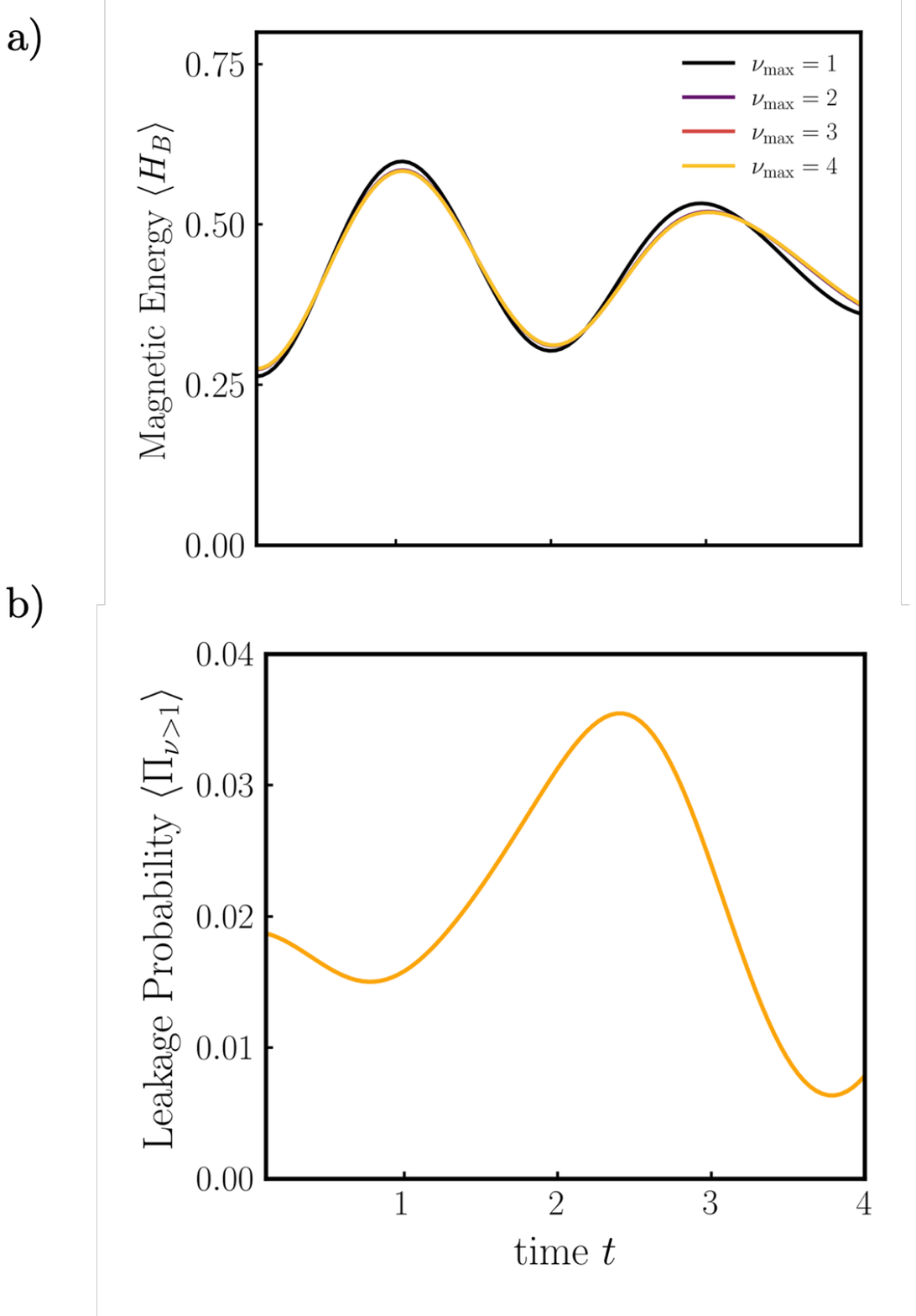}
    \caption{\textit{Accuracy of the $\nu$ Truncation} \textbf{a)} Calculation of the time dependent magnetic energy for the low-energy state $\ket{5}$, defined in Eq.~\eqref{eq:low_energy_state}, as the maximum $\nu$ value is increased for $g=0.5$ and $N_{\omega}=10$. This choice of digitization resolution ensures that any errors due to the $\omega_1,\omega_2$ truncation will be at the percent level. \textbf{b)} The probability $\langle\Pi_{\nu>1}\rangle$ of leaking out of the $\nu=0,1$ sector as a function of time for $g=0.5, N_{\omega}=10$ and $\nu_\text{max} = 5$.}
    \label{fig:nu_truncation}
\end{figure}

In addition to the truncation of the gauge field, the discrete quantum number $\nu$ must also be truncated. An advantage of the mixed basis is that the low-lying eigenstates are supported solely in the $\nu=0,1$ sector, and so only a single qubit is required to represent this quantum number. Furthermore, $\htp$ only connects states that have $|\nu-\nu'|\leq 1$, and so for low-energy simulations heavy truncations on this degree of freedom will have little effect on the precision. To study the effect of different $\nu$ truncations on the time evolution of observables, the system is initialized in the state
\begin{align}\label{eq:low_energy_state}
\ket{5}=\frac{1}{\sqrt{5}}(\ket{E_0}+\ket{E_1}+\ket{E_2}+\ket{E_3}+\ket{E_4}) \, \,
\end{align}
where $\ket{E_n}$ is the $n^{th}$ eigenstate of the digitized Hamiltonian and then time evolved according to different maximum $\nu$ values $\nu_{\text{max}}$. Fig.~\SubFigRef{fig:nu_truncation}{a} shows the time dependence of the expectation value for the magnetic part of the Hamiltonian $H_B$. As $\nu_{\text{max}}$ is increased, the values of $\langle H_B\rangle$ rapidly converge to the same curve, suggesting that $\nu_{\text{max}}=2$ is already sufficient for extracting high-precision estimates of observables.

The leakage out of the $\nu=0,1$ sectors is also studied. Defining the projector onto states with $\nu > 1$ as
\begin{equation}
    \Pi_{\nu>1} = \mathbb{I} - \sum_{J_1,J_2}\ket{J_1J_20}\bra{J_1J_20} + \ket{J_1J_21}\bra{J_1J_21},
\end{equation}
the expectation value $\langle \Pi_{\nu > 1}\rangle$ encodes the probability that the state has leaked out of the low-energy $\nu=0,1$ subspace. Fig.~\SubFigRef{fig:nu_truncation}{b} shows the leakage probability as a function of time. As the state evolves, the leakage does not grow noticeably over time, showing that a low-energy state with little support on values of $\nu > 1$ will not evolve to states with larger support over a range of times.

\section{Circuits for Time Evolution}
\label{sec:MixedBasisCircuits}

In this section the digitization presented above is translated into quantum circuits that can be implemented on current quantum devices. The circuit architecture for implementing time evolution is not unique, and two distinct approaches are developed that have different depths based on the level of digitization that is required for the simulation task. 

In the first approach to circuit design, a single Trotter step is formed from component sub-circuits. This has the advantage of avoiding any pre-computation of circuits as  the form of the components are already known. Additionally, most of these components have an asymptotic scaling that is no more than quadratic in $n_q$. For the component implementing a diagonal unitary that encodes various transcendental functions, exact implementations scale exponentially in the number of elementary gates with $n_q$, the number of qubits used to digitize $\omega_1$ and $\omega_2$. While this is seen as an unavoidable feature of this system, implementing such gates up to a desired level of precision scales more favorably than the worst-case exponential~\cite{Li:2024lrl}. 

This approach gives the greatest benefit when high resolution in field space is required. When the number of qubits per plaquette is of order one, it is advantageous to employ a different approach. This begins by directly decomposing the Hamiltonian into a sum of Pauli strings and implementing the time evolution operator by a series of rotations. This approach scales exponentially in $n_q$, although dropping strings with weights below some threshold is found to give circuit-depth reductions that outweigh the losses in accuracy. The second approach is more suited to simulation tasks where only low resolution is field space is necessary. The following sections give explicit circuit decompositions for each of these approaches and studies the accuracy and different approximations that go into each. The relative resource counts are then compared.

\subsection{Circuit from Differential Operators}
\label{subsec:DifOpCircuits}

\begin{figure*}[t]
    \centering
\includegraphics[scale=.65,trim = {0 180 0 180}]{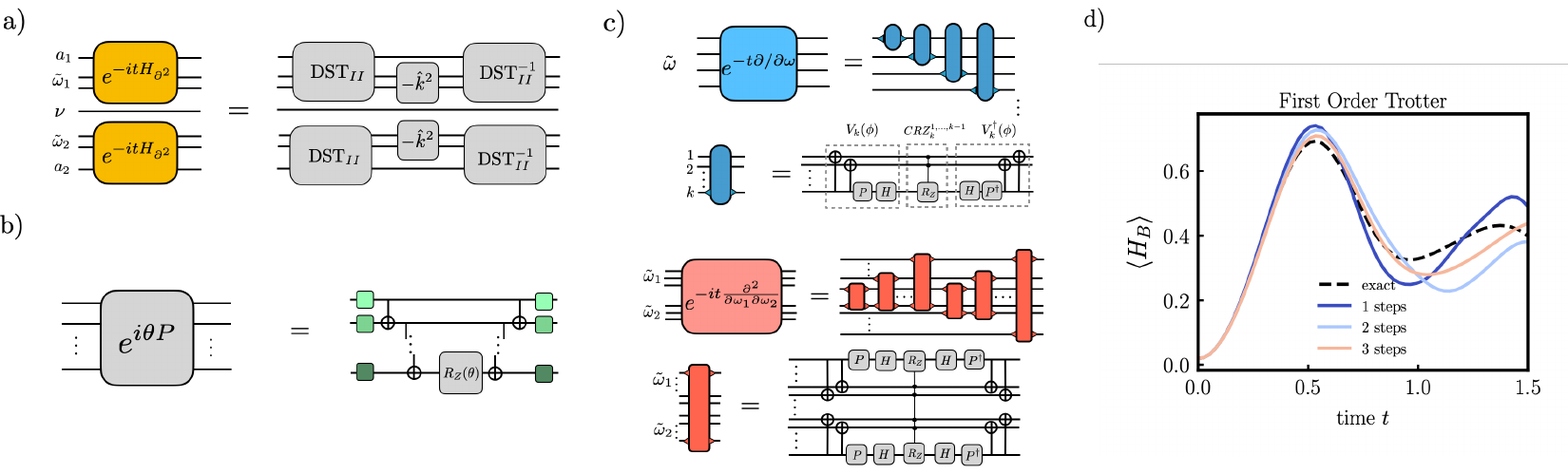}
    \caption{\textit{Circuits for the  Digitization Presented in Sec.~\ref{subsec:DifOpCircuits}, which constructs sub-components for each class of differential operators} \textbf{a)} Implementation of the exact exponential of the second derivative, making use of Discrete Sine Transforms. This subcircuit only entangles qubits within each $\omega$ register and does not tuch the $\nu$ register. The ancillas $a_1,a_2$ associated with each $\omega$ register are the outermost qubits. \textbf{b)} An important circuit identity used in this work: The exponential of an $N-$body Pauli string $P$ is implemented by the staircase of $CX$s connecting all the qubits participating in the gate, followed by an $R_Z(\theta)$ rotation on one of the qubits and then another $CX$ staircase. The green single qubit gates denote a local change of bases to change $P$ into a string of $Z$s on all qubits e.g. if there is an $X$ on qubit $i$ then this will be a Hadamard $H$ and if it is a $Y$ then it will be a Hadamard followed by a phase gate $HS$. The different shades denote that the local basis change can depend on the particular Pauli string $P$ being implemented.
   \textbf{c)} Circuits for implementing first order derivatives and mixed derivatives using the finite difference representation. \textit{(top)} Implements the exponential of a single first derivative.Each element in this circuit implements one of the terms in the sum Eq.~\eqref{eq:partial_MPO}. The bottom qubit of each step is the target qubit for both the $V_k$ and $CRZ$ gates. \textit{bottom} An example of one of circuit elements $e^{-it\frac{\partial}{\partial\omega_1}\frac{\partial}{\partial\omega_2}}$ that interleaves the top circuit. \textbf{d)} Noiseless simulator results for the expectation value of the magnetic Hamiltonian at a different number of Trotter steps.}
    \label{fig:fine_dig_circ}
\end{figure*}

This approach proceeds by organizing the terms in the
two-plaquette Hamiltonian according to the highest order
differential operator each term contains. In particular, the terms in the Hamiltonian are grouped in the following way
\begin{equation}
    \htp = H_{\partial^2}+H_{\partial}+H_{\omega}
\end{equation}
where $H_{\partial^2}$ collects all the terms that have only a second derivative, $H_{\partial}$ contains any terms that have a first order derivative, and $H_{\omega}$ is all remaining terms, \textit{i.e.} those that are only functions of $\omega$ with $\nu$-dependent coefficients. The total time evolution operator is Trotterized according to these groupings, $e^{-it\htp}\approx e^{-itH_{\partial^2}}e^{-itH_{\partial}}e^{-itH_{\omega}}$.

The circuits for $e^{-itH_{\partial^2}}$ and $e^{-itH_{\omega}}$ are straightforward, and mirror the circuits involved in simulating the scalar field. The Laplacian term is implemented by switching to the Fourier basis with a DST$_{II}$, which is based off of the Quantum Fourier Transform (QFT)~\cite{klappenecker2001discrete}. In this basis each second derivative is diagonal and proportional to the square of the conjugate momentum operator $\hat{k}^2$,
\begin{equation}
    \hat{k}^2= \frac{\pi}{\omega_{\text{max}}}\left(\frac{2^{n_q}-1}{2}I-\sum_j^{n_q-1}2^{j-1}Z_j\right)^2
\end{equation}
The circuit implementing a DST$_{II}$ is shown in Fig~\SubFigRef{fig:fine_dig_circ}{a}. While using the QFT as an algorithmic primitive, extra elements (not shown) are required to enforce the boundary conditions. The DST circuit requires an additional ancilla qubit changing the qubit count per gauge variable to be $n_{q}+1$. $H_{\omega}$ is completely diagonal in the mixed basis and so it can be represented as sum of Pauli $Z$ strings. It is implemented by a series of $R_Z{(\theta)}$ with $\nu$-dependent rotation angles, as shown in Fig.~\SubFigRef{fig:fine_dig_circ}{b}.

The construction of the circuits for $e^{-itH_{\partial}}$ is more involved. Recalling the results from Sec.~\ref{subsub_sec:groupel_basis} and Sec.~\ref{subsub_sec:mixed_basis}, the matrix representation of the Hamiltonian is organized into $(\nu,\nu')$ sectors, where nontrivial sectors satisfy the selection rule $|\nu-\nu'|\leq 1$. The $\nu=\nu'$ case only involves the circuits already discussed. For $\nu\neq \nu'$, the Hamiltonian $H_{\partial}$ is expressed as
\begin{equation}
    H_{\partial} = \sum_{\nu}H^{+}_{\nu}\ket{\nu}\bra{\nu+1}+\sum_{\nu}H^{-}_{\nu}\ket{\nu}\bra{\nu-1}
\end{equation}
where the expressions for $H^\pm_\nu$ are given in Appendix~\ref{app:MB}.  

The coupling of different $\nu$ sectors is reminiscent of a finite difference matrix in $\nu$ space. As already noted, implementing such an operator exactly requires a prohibitive number of gates. This is where the expressive power of the mixed basis shines through. Truncating to $\nu_{max}=1$ gives the off-diagonal Hamiltonian the form
\begin{align}
H_{\partial}=\sigma^{+}H^{+}_{0}+\sigma^{-}H^{-}_{1}
\end{align}
where $\sigma^{\pm}$, defined  as
\begin{align}
\sigma^{\pm}=\frac{1}{2}(X\pm iY) \, ,
\end{align}
acts on the single-qubit $\nu$-register. Applying $HS^{\dagger}$ to the $\nu$-register turns the projectors into 
\begin{align}
\sigma^{\pm}\mapsto \frac{1}{2}\left(Z\pm iX\right) \,\, ,
\end{align} 
where $H$ is a Hadamard and $S$ is a $\pi/2$ phase gate. Thus under this transformation, $H_{\partial}$ maps to
\begin{align}
H_{\partial}\mapsto \frac{1}{2}\left(Z\otimes H_{\text{sym}}-iX\otimes H_{\text{anti}}\right)
\end{align}
where $H_{\text{sym}}, H_{\text{anti}}$ are the symmetric and anti-symmetric parts of $H^{+}_0$,\footnote{$H_0^{+}$ and $H_1^{-}$ are hermitian conjugates}
\begin{align}
H_{\text{sym}}&=H_0^{+}+H_1^{-}\nonumber \\
H_{\text{anti}}&=H_0^{+}-H_1^{-} \, .
\end{align}
Explicitly, these components are given by
\begin{align}
    H_{\text{sym}} &= \frac{1}{\sqrt{3}}\pdv{}{\omega_1}\pdv{}{\omega_2}+\frac{1}{4\sqrt{3}}\cot \frac{\omega_1}{2}\cot \frac{\omega_2}{2}\nonumber\\
    H_{\text{anti}} &= -\frac{1}{2\sqrt{3}}\left(\cot \frac{\omega_1}{2}\pdv{}{\omega_2}+\cot \frac{\omega_2}{2}\pdv{}{\omega_1}\right)\nonumber \,\,.
\end{align}
The complete time evolution induced by the first derivatives may be approximated by
\begin{align}\label{eq:h_partial_evo}
    e^{-itH_{\partial}} &\approx e^{-itg^2Z\otimes H_{\text{sym}}}e^{-itg^2X\otimes H_{\text{anti}}}\nonumber\\
    &\approx e^{-i\frac{g^2t}{\sqrt{3}}Z\otimes\pdv{}{\omega_1}\pdv{}{\omega_2}}e^{-i\frac{g^2t}{4\sqrt{3}}Z\otimes \cot \frac{\omega_1}{2}\cot \frac{\omega_2}{2}}\nonumber\\
    &\hspace{30pt}e^{-\frac{g^2t}{2\sqrt{3}}X\otimes \cot \frac{\omega_1}{2}\pdv{}{\omega_2}}e^{-\frac{g^2t}{2\sqrt{3}}X\otimes \cot \frac{\omega_2}{2}\pdv{}{\omega_1}}\nonumber\\
\end{align}
where the tensor product separates the $\nu$ register from the $\omega$ registers and the dependence of gauge coupling has been reinserted.

Therefore, the problem of implementing the time evolution due to the $H_{\partial}$ terms is reduced to exponentiation of the finite difference operator. It has been recognized that this operator can be recast as a sum of Matrix Product Operators (MPO)~\cite{sato2024hamiltonian} with an induced Trotter error of $t^2(n_q-1)/2$ . To see this, one starts from the identity
\begin{align}\label{eq:partial_MPO}
\pdv{}{\omega} &\mapsto\frac{-i}{2\delta\omega}\sum_k I^{\otimes n_{q}-k}\otimes \\
    &V_k\left(-\frac{\pi}{2}\right)\left(Z_k\otimes\ket{1}\bra{1}^{\otimes k-1}\right)V_k^{\dagger}\left(-\frac{\pi}{2}\right) \,. \nonumber
\end{align}
where $\hat Z_k$ is the Pauli $\hat Z$ operator on the $k^{th}$ qubit in the corresponding register and $V_k(\phi)$ is a unitary change of basis defined by
\begin{equation}
    V_k(\phi) = \left(\prod_m^{k-1}\text{CX}_k^m\right)P_k(\phi)H_k \,
\end{equation}
where $\text{CX}_k^m$ is a CNOT gate with control on qubit $m$ and target on qubit $k$, $P_k(\phi)=\text{diag}(1,e^{i\phi})$ is a phase gate on qubit $k$, and $H_k$ is a Hadamard on qubit $k$, not to be confused with a Hamiltonian. After taking the exponential and Trotterizing the sum over $k$ in Eq.~\eqref{eq:partial_MPO}, the operator $e^{-t\partial/\partial\omega}$ can be implemented by a sequence of controlled $CRZ_k^{1,\dots,k-1}(t/\delta\omega)$ rotations for each qubit in the register $k$, where $CRZ_k^{1,\dots,k-1}(\theta)$ is a controlled $R_Z(\theta)$ with target $k$ and controls $1,\dots,k-1$.

Inspection of Eq.~\eqref{eq:h_partial_evo} shows that it is insufficient to implement terms just of the form $e^{-t\partial/\partial\omega}$, as there are entangling operations between the two gauge variable registers and the $\nu$ register. Therefore, it is necessary to modify the above circuits by interleaving the extra operations between each set of $V_k$ and $V_k^{\dagger}$. As an example, the first term will need to implement an operator which is the product of a first derivative on each $\omega$. Expanding the MPO expression for the first derivative gives
\begin{align*}
    \pdv{}{\omega_1}\pdv{}{\omega_2} &\mapsto -\frac{1}{4\delta\omega^2}\sum_{k,k'}I^{\otimes n_q-k}I'^{\otimes n_q-k'}\otimes\\
    &\hspace{10pt}V_kV'_k\left(Z_kZ_k'\otimes \ket{1}\bra{1}^{\otimes k-1}\ket{1'}\bra{1'}^{\otimes k'-1}\right)V_k^{\dagger}V_k'^{\dagger}
\end{align*}
where the (un)primed operators denote operators acting on the ($\omega_1$)$\omega_2$ register and the argument of the $V_k,V'_k$ have been suppressed. A similar interleaving can be made for the remaining terms in Eq.~\eqref{eq:h_partial_evo}, and for concreteness, the circuits structures for the case just discussed is shown in Fig.~\SubFigRef{fig:fine_dig_circ}{c}.

Fig.~\SubFigRef{fig:fine_dig_circ}{d} shows different numbers of first-order Trotter steps of these circuits at various times $t$. If the noise profile of the device is partially or completely characterized, there exists an optimal step size for a target total simulation time~\cite{Clinton:2021pdy,Zemlevskiy:2023eyw}. For even a single first-order step, this method remains accurate at up to a step size of approximately $t=0.5$, at which point dephasing is observed and deviations from the exact curve begin to grow. Modifying these circuits to perform a second-order step, with error that falls off as $\mathcal{O}(t^3)$ is straightforward, and requires modifications of the $V_k$ circuits as outlined in~\cite{sato2024hamiltonian}.

\subsection{Circuits by Pauli Truncation}
\label{subsec:coarse}

Simulating time evolution requires converting the digitized time evolution operator $U=e^{-i t \htp}$ into a sequence of two-qubit gates directly implementable on a quantum device. The most straightforward way to do this, while not the most efficient, is to calculate $\mathcal{D}(\htp)$. Once this decomposition is performed, the sum in the exponent of $U$ can be split into a product over Paulis using a Trotter-Suzuki formula, which approximates the exact time evolution operator. This time evolution operator $U_{T}(t)=\prod_Pe^{-i t c_P P}$, where $(P,c_p)\in \mathcal{D}(\htp)$, can be turned into circuit-level operations using the identity displayed in Fig.~\SubFigRef{fig:fine_dig_circ}{b}. Generally this procedure will have a two-qubit gate count that scales as $\mathcal{O}(4^{2n_q+n_{\nu}})$. However, many of the rotations, particularly those corresponding to high weight strings, will have small angles that do not greatly impact the expectation value of observables over some range of evolution times. The rotations with small angles less than $\theta_{\text{min}}$ can be dropped, which gives a compressed approximation of the exact time evolution operator.

It is necessary to understand the impact of dropping small angles for different levels of gauge field digitization and at different values of the gauge coupling. By removing these terms, the system is evolved instead by the Hamiltonian 
\begin{align}
H_{\delta}&=\htp + \theta_{\text{min}}V \nonumber \\
V&=-\sum_{c_P\leq \theta{\text{min}}}\left(\frac{c_P}{\theta_{\text{min}}}\right)P
\end{align}
where $\delta$ corresponds to the percentage of terms in the Pauli decomposition that have coefficient below $\theta_{\text{min}}$. The time evolution implemented on the device is given by the unitary 
\begin{align}
U_{\delta}(t)=UU_{\theta}(t)
\end{align}
where the $U_\theta$ (and its component) matrix, given by
\begin{align}
U_{\theta}(t)&=e^{-i\theta_{\text{min}}\int_0^td\tau V_I(\tau)} \nonumber \\
V_I(\tau)&=e^{-i\tau\htp} Ve^{i\tau\htp} \,\, ,
\end{align}
is the error term in the interaction picture. Therefore the extent to which the perturbation $V$ grows under the dynamics of the two plaquette system directly controls the appropriate level of truncation. The error in evolution on some observable $O$ is quantified by
\begin{align}
\langle O(t) \rangle_{\delta}=\langle U^{\dagger}_{\theta} O(t)U^{\phantom{\dagger}}_{\theta} \rangle \, .
\end{align}
which is the time-evolved value of the observable. This is compared to the untruncated value $\langle O(t) \rangle$ ($\delta=0$).

\begin{figure*}[t]
    \centering
    \includegraphics[scale=.65, trim = 0 150 0 130]{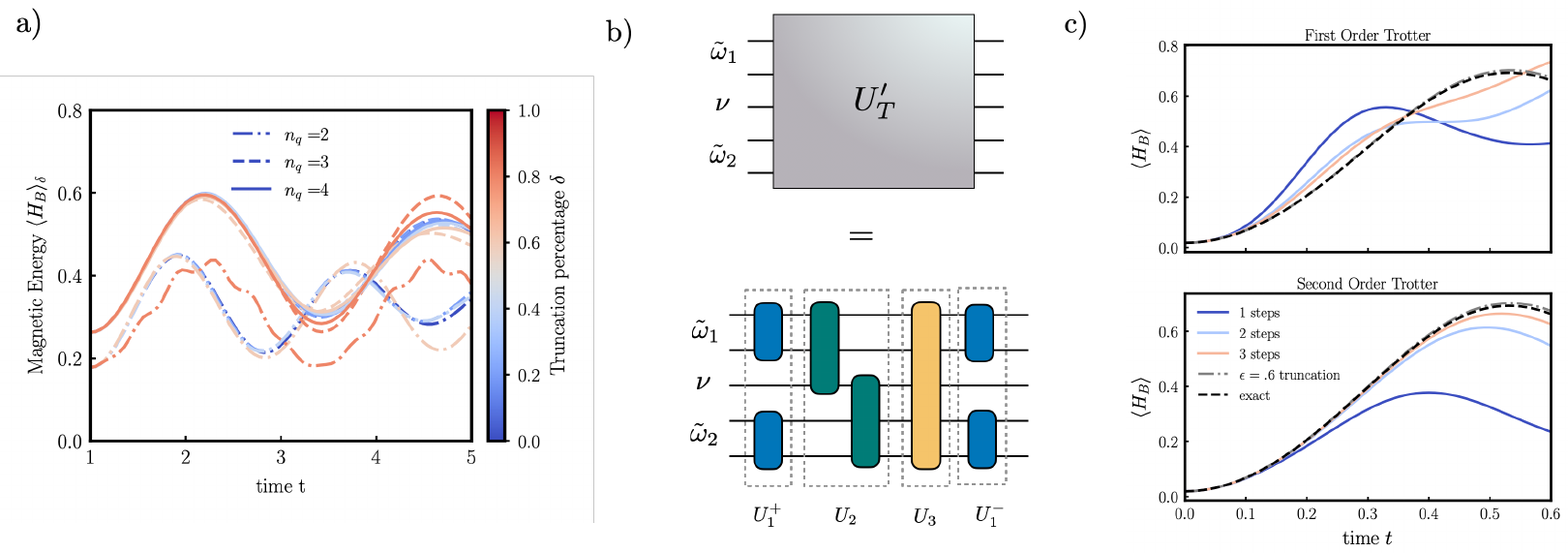}
    \caption{\textit{Approximation for Time Evolution Circuits Presented in Sec.~\ref{subsec:coarse}} \textbf{a)} The expectation value of the magnetic part of the Hamiltonian for differing digitization resolutions and percentage of terms dropped. Rapid convergence is observed for $n_q\geq 3$ and for truncation levels $\delta<.6$. \textbf{b)} Different stages of the time evolution operator for a second-order Trotter step. The silver circuit denotes the entire Trotter step. The blue sub-circuits acts only on the $\omega$ registers, typically involving the largest angles, and is identical for each. The green sub-circuits couple each $\omega$ register to the $\nu$ register and the circuit structure is again identical for both $\omega$. The yellow sub-circuit couple everything and generally contain the smallest rotations, therefor being the elements most affected by small-angle truncations. It also involves the highest weight terms, and so focusing on truncating these elements allows the circuit to be compressed at only a small cost in algorithmic error. \textbf{c)} The accuracy of first- and second-order Trotter on magnetic observables. Repeating the blue sub-circuit in the second step allows the circuit to take much larger time steps while only needing a small amount of extra gates to implement.}
    \label{fig:circuit_stuff}
\end{figure*}

To understand the effect of Pauli truncation error, the system is initialized in the state $\ket{5}$ defined in Eq.~\eqref{eq:low_energy_state} and the expectation value of the magnetic part of the Hamiltonian $H_B$ is calculated. Fig~\SubFigRef{fig:circuit_stuff}{a} shows this expectation value for various levels of truncation and gauge field digitization. For a coarse field digitization of $n_q=2$, the expectation value is generally more sensitive to truncations. As the digitization resolution is increased, a larger percentage of the terms in the Pauli decomposition are able to be dropped, showing that an increasingly small fraction of the terms in the decomposition are responsible for capturing the relevant dynamics. For diagonal matrices, this can be understood via the Sequency Hierarchy  Truncation~\cite{Li:2024lrl}. The extension to non-diagonal matrices has not yet been developed, but similar arguments should hold. 

Having truncated the Hamiltonian to retain only the Pauli strings that are most relevant for implementing time evolution, the time evolution operator $U_{\delta}$ must be converted into a quantum circuit. To minimize the number of CNOTs each Trotter step contains, the Pauli rotations are grouped to model the terms that appear in the Hamiltonian. This structure is presented in Fig.~\SubFigRef{fig:circuit_stuff}{b}. The first group includes strings that only act on individual $\omega$ registers, the second group involves strings that couple a single $\omega$ register to the $\nu$ register and have a $Z$ acting on the $\nu$ register, and the third group includes strings that act on all of the qubits. The reason for classifying the rotations in this way are two-fold. First, as already mentioned, this grouping respects how the digitized Hamiltonian couples different quantum numbers, and so this structure establishes a hierarchy in the significance of how each term contributes to time evolution. The terms in group three typically have much smaller rotation angles and are the ones to be dropped first. Second, clustering terms that acts on smaller subsets of qubits makes compilation easier, as only CNOT cancellations over a subset of qubits need to be considered. For compilation the open source software package $\textbf{bqskit}$ is used~\cite{bqskit}. The circuit that implements the $i^{th}$ group is defined to be $U_i$, and so the first-order product formula is $U_T=U_3U_2U_1$. As group one is restricted to weight-$n_q$ Paulis, it will generally involve less CNOTs to implement, and therefore it is advantageous to use the second-order Trotter formula $U'_T=U^-_1U_3U_2U^+_1$, where $U_1^{\pm}$ involve the same set of operations as $U_1$ with the rotation angle $c_P t$ replaced by $c_P t/2$ and $\pm$ denotes the same/reversed ordering of operations as compared to $U_1$. The explicit rotations for each term $U_1,U_2,U_3$ are recorded in Appendix~\ref{sec:appendixA}.

Fig.~\SubFigRef{fig:circuit_stuff}{c} shows the evolution of the magnetic energy using both first- and second-order formulas for a variable number of Trotter steps. The second-order method allows for longer steps, giving similar levels of accuracy as three steps of first-order Trotter but a notably smaller amount of CNOTs. Also shown for comparison is the exact evolution at $\delta=0.66$ levels of truncation. Even with over half of the Pauli strings dropped, for long enough times, the error will be subleading to algorithmic error due to Trotterization.

\subsection{Resource Estimates}
To count the gates needed to implement a Trotter step using the differential operator method, the three stages are considered individually. For the Laplacian term, the quantum DST circuits can be implemented using $\mathcal{O}(n_q^2)$ elementary quantum gates, where the quadratic scaling is a direct result of the use of the Quantum Fourier Transform. The diagonal momentum operator requires an additional $n_q^2$ gates, so this entire stage requires only $\mathcal{O}(n_q^2)$.

Implementing the time evolution $e^{-t\partial/\partial\omega}$ requires at most $9n_q^2-33n_q+34$ CNOT gates for $n_q\geq 3$~\cite{sato2024hamiltonian}. Using the interleaving method previously discussed, the two qubit gate count is the product of the gate counts for the two types of terms that are interleaved, e.g. $e^{-it\pdv{}{\omega_1}\pdv{}{\omega_2}}$ will require $(9n_q^2-33n_q+34)^2\sim\mathcal{O}(n_q^4)$. The decomposition of the other terms involving functions of $\omega_1,\omega_2$ in Eq.~\eqref{eq:h_partial_evo} will asymptotically require $\mathcal{O}(2^{2n_q+1})$ $Z$ strings that must be classically pre-computed. This situation mirrors the counts estimated for compact \uone gauge theories~\cite{Kane:2022ejm}. In this case it was found that using the Walsh basis to encode functions of the gauge variables and truncating interactions with small angles was sufficient to break the exponential scaling to polynomial, suggesting an efficient path forward to completely reducing the scaling to polynomial in $n_q$. In absence of these more specialized methods the gate count for a single Trotter step using the differential operator method will scale as $\mathcal{O}(2^{2n_q+1})$. It is once more emphasized that the only place where the exponential scaling enters is in exactly implementing the diagonal functions of $\omega_1,\omega_2$.

Turning now to the gate counts for the Pauli truncation method, even though a large percentage of the gates can be dropped to calculate time-evolved quantities up to some specified level of error, the scaling will generally be exponential in the number of qubits per plaquette. However, for large-scale simulations it may be more efficient to utilize the qubit budget for increasing the lattice volumes as opposed to increasing the resolution in field space. For instance, on a two-dimensional $(L+1)\times(L+1)$ lattice, the dominant interaction is between pairs of plaquettes. The number of such interactions is $\binom{L^2}{2}$, which is quartic in the linear dimension $L$. This suggests that the only exponential scaling is due to resolving the Hilbert space at each physical link. As already seen in the previous section, ground state properties can be resolved with precision of $10^{-3}$ already at $n_q=3$. Understanding the resource requirements for the Pauli truncation circuits up to per-mille precision is done by considering the number of two-qubit gates needed for time-evolution.

\begin{figure}
    \centering
    \includegraphics[scale=0.3]{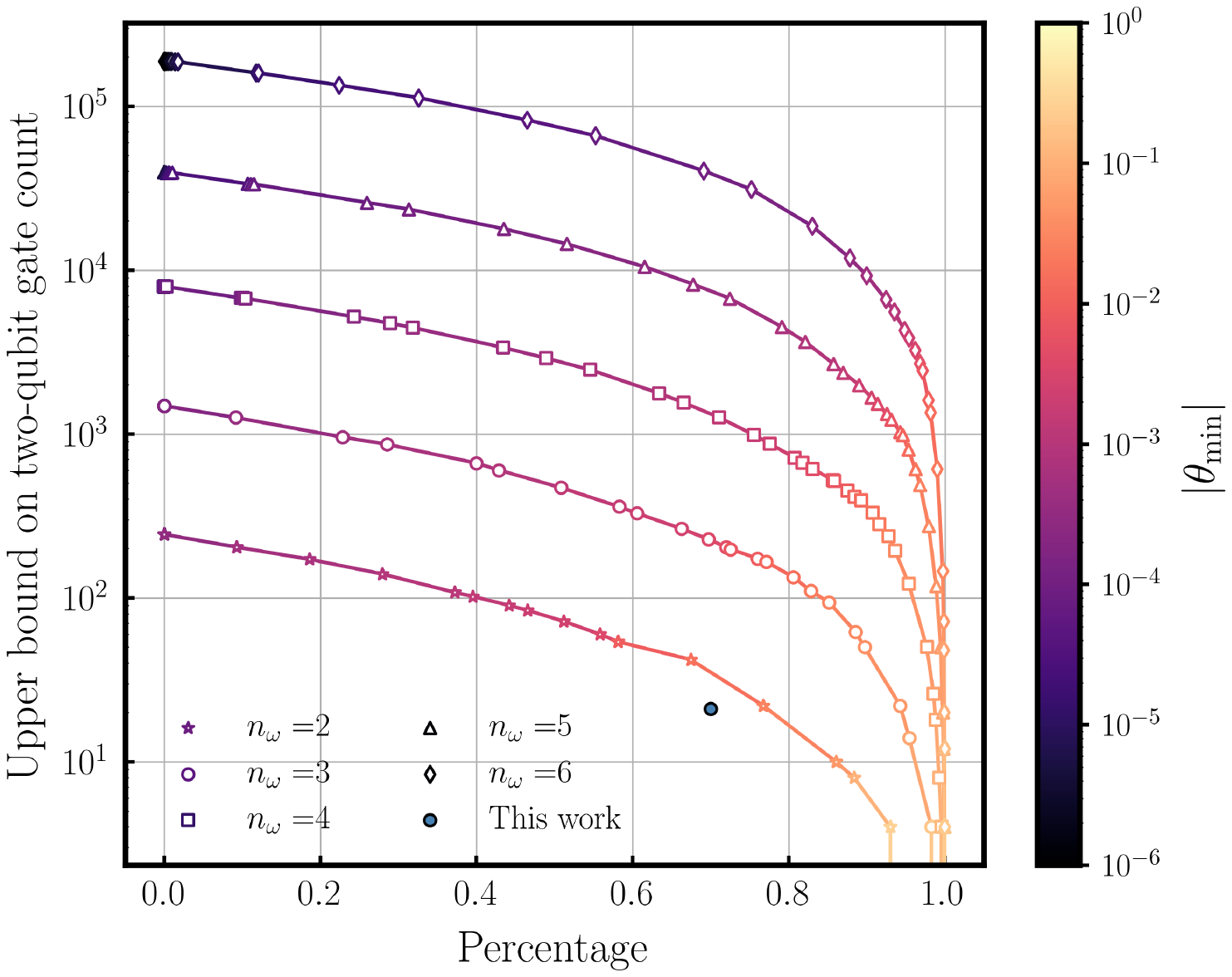}
    \caption{\textit{Upper Bounds on Resource Estimates for the Pauli Truncation Circuits} The cumulative weight of the Pauli decomposition of $\htp$ is shown as a function of the truncation percentage $\delta$. The color of the curves denotes the magnitude of the coefficient in the decomposition. The blue point demonstrates the truncation level and actual gate depth used in the quantum simulation in this work, highlighting the effect of circuit compilation.}
    \label{fig:gate_count}
\end{figure}

Analytically decomposing $\htp$ into Pauli strings for arbitrary $n_q$ is challenging, and so an upper-bound on the number of two-qubit gates is estimated by observing that every Pauli string of weight $w\geq 2$ in the Hamiltonian will require $2(w-1)$ CNOTs, as shown in Fig.~\SubFigRef{fig:fine_dig_circ}{b}. The estimate
\begin{equation}
    \#(\text{Two Qubit}) \leq \sum_{P\in\mathcal{D}(\htp)}2(w(P) - 1)
\end{equation}
is numerically calculated for digitization resolutions of $\htp$ up to $n_q=6$. Fig.~\ref{fig:gate_count} shows the upper bound as a function of the percentage of truncated strings from the Hamiltonian. This upper bound still scales exponentially in the number of qubits per plaquette, although this scaling can be improved by a factor of $2-3$ with appropriate truncation. In addition, careful compilation of the circuits can bring this number down by another factor of $\sim 2$. However, compilation is also generally a hard problem, so to compare the raw estimate to the number of post-compilation gates, the number of gates used on the quantum device in this work is also shown.

\section{Real-Time Dynamics in Two Plaquettes}
\label{sec:quantum_results}
\begin{figure}
     \centering
     \includegraphics[scale=.4]{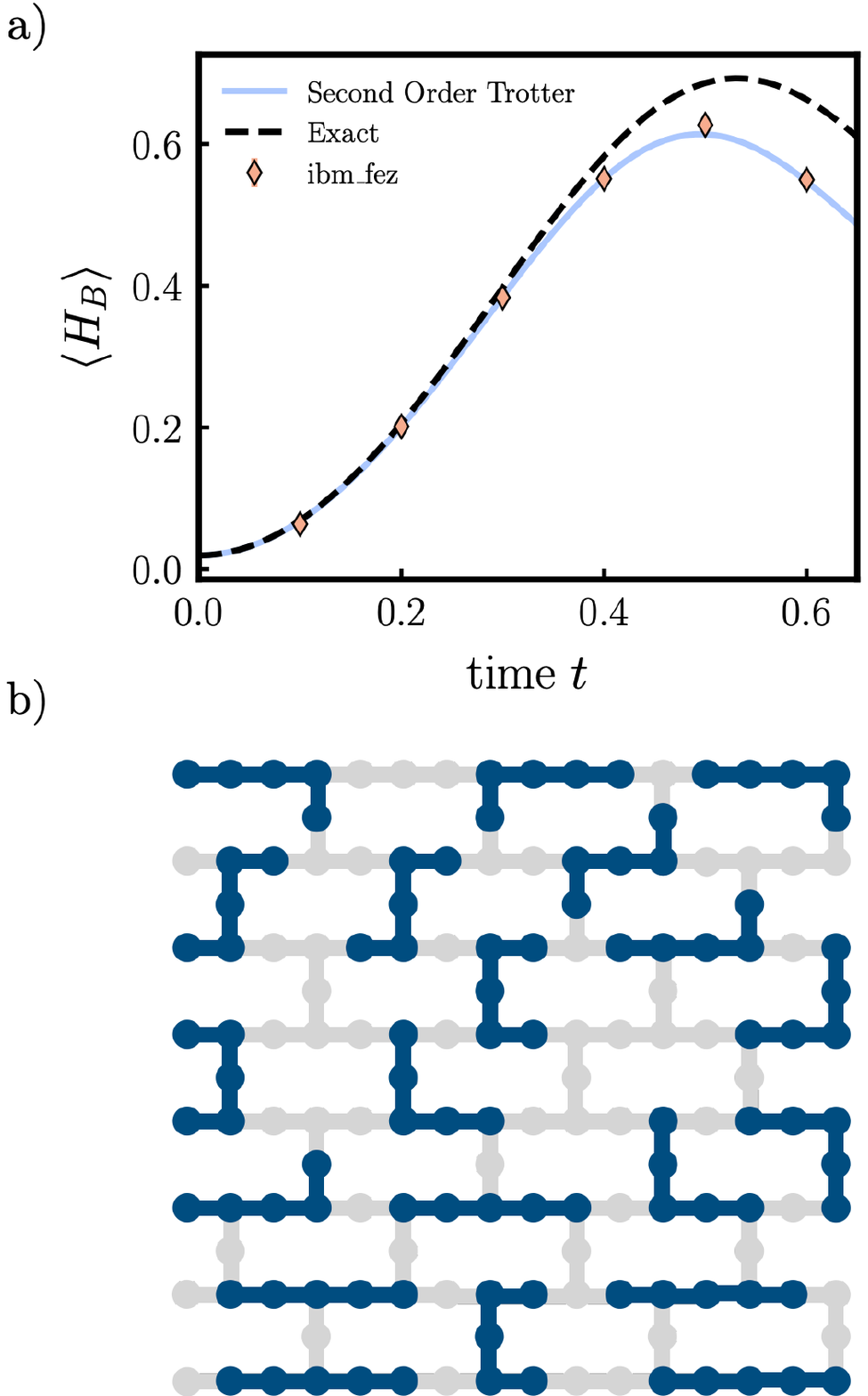}
     \caption{\textit{Quantum Simulation of $\langle H_B\rangle$ Dynamics} \textbf{a)} The time evolution due to $\htp$ at gauge coupling $g=0.5$. The dashed line corresponds to the solution obtained by directly taking the matrix exponential of the digitized Hamiltonian and the solid line is the Trotterized evolution. The points are the mitigated results from $\textbf{ibm\_fez}$. \textbf{b)} The physical layout of the Heron chip. The highlighted regions are an example of the parallel layouts used to generate the data point at $t=0.6$.}\label{fig:quantum_results}
 \end{figure}

This section presents the results of a real-time simulation on the 156 qubit $\textbf{ibm\_fez}$ superconducting quantum chip. The circuit architecture used is the Pauli truncation approach presented in Sec.~\ref{subsec:coarse} at a truncation level of $\delta=0.66$ on five qubits ($n_q=2,n_{\nu}=1$). Preparing the example low-energy state $\ket{5}$ in Eq.~\eqref{eq:low_energy_state} requires additional circuits for time evolution. To isolate the performance of the time evolution circuits, the system is initialized in the product state $\ket{00000}$. Note that all of the states in the Hilbert space that $\htp$ acts upon satisfy Gauss' law, and so this product state is still a valid physical state. The system is then time evolved at gauge coupling $g=0.5$ and the expectation values of the magnetic part of the Hamiltonian $\langle H_B\rangle$ are measured. For the gauge coupling considered here, $\mathcal{D}(H_B)$ has the following explicit decomposition into Pauli observables:
\begin{align}
H_B &= 0.386\; I\otimes I\otimes I\otimes I\otimes I \nonumber \\
&\hspace{10pt}-0.0701\;I\otimes I \otimes I\otimes I\otimes Z \nonumber \\
&\hspace{20pt} -0.1445\;I\otimes I\otimes I\otimes Z\otimes I\nonumber\\
&\hspace{30pt}+ 0.0326\;I\otimes I\otimes I\otimes Z\otimes Z \nonumber \\
&\hspace{40pt} -0.0701\; I\otimes I\otimes Z\otimes I\otimes I\nonumber \\
&\hspace{50pt}-0.1448\; I\otimes Z\otimes I\otimes I\otimes I \nonumber \\
&\hspace{60pt} + 0.0325\;I\otimes Z\otimes Z\otimes I\otimes I  \, .
\end{align}
Fig.~\SubFigRef{fig:quantum_results}{a} shows the evolution of this observable from time $t = 0.1$ to $t=0.6$, in increments of $0.1$. To make maximum use of the device, up to 20 copies of the circuit were run in parallel at different locations on the chip, with 100 active qubits for every job. To limit the effect of cross-talk error between different parallel circuits, the layout routing was chosen such that at least one inactive qubit separated any active qubits in different subcircuits. An example of the layout used to generate the $t=0.6$ data point is shown in Fig.~\SubFigRef{fig:quantum_results}{b}.

Due to the inherent noisiness of current machines, utilizing various layers of error mitigation was crucial in obtaining these results. Once the bare circuits are created and layouts chosen, Dynamical Decoupling (DD)~\cite{Viola:1998jx,Ezzell:2022uat} was added to suppress idle errors and further mitigate the effects of cross-talk between qubits. In addition, Pauli Twirling (PT)~\cite{Wallman:2015uzh} and Twirled Readout Error eXtinction (TREX)~\cite{Berg:2020ibi} are applied to each circuit. In many cases, these methods allows the dominant noise channels of the device to be approximated as a depolarizing channel. The benefit is that the impact of a depolarizing channel on observable estimates is effectively dealt with using the final mitigation technique, Operator Decoherence Renormalization (ODR)~\cite{Farrell:2023fgd,Farrell:2024fit,Urbanek:2021oej,ARahman:2022tkr}.

ODR is a very powerful error mitigation technique that allows
the raw data from the device to be `renormalized' in order to extract accurate physical results. For this method, two types of circuits are run on the quantum device. The first is the circuit for the physical process of interest, referred to as the physics circuit. The second, called the mitigation circuit, is a circuit that has the same pattern of single- and two-qubit gates, but is classically efficient to simulate. Classically simulating the mitigation circuit provides a noise-free and therefore numerically exact result to which the quantum simulation can be compared.  Assuming the dominant error channel can be approximated by sufficiently weak depolarizing noise, the quantum estimate of the observable $O$ will deviate from the true value by a multiplicative factor $p_O$, \textit{i.e.} $\langle O\rangle_{\text{meas}} = p_O \langle O\rangle_{\text{true}}$. If the physics circuit is affected by the noise in the same way as the mitigation circuit, the factors of $p_O$ between both are the same. Therefore, the renormalized estimate of the physical observable is given by
 \begin{equation}
     \langle O_{\text{phys}}\rangle_{\text{true}} = \langle O_{\text{phys}}\rangle_{\text{meas}}\frac{\langle O_{\text{mit}}\rangle_{\text{true}}}{\langle O_{\text{mit}}\rangle_{\text{meas}}}
 \end{equation}
 where the subscripts $\textit{meas} /\textit{true}$ denote the estimate using the raw/renormalized quantum data and the subscripts $\textit{phys} /\textit{mit}$ denote the physics/mitigation circuits.

 For this technique to work, the choice of mitigation circuit is of central importance. For the smaller time steps, the physics circuits were `Cliffordized', meaning that the rotation angles of all $R_Z$ gates were rounded to the nearest $\pi / 2$. This transforms the mitigation circuit into a Clifford circuit with the same gate instructions as the physics circuit. ODR is inapplicable for long enough time steps, as the rounding procedure returns a circuit that maps to a state that yields $\langle O_{\text{mit}}\rangle_{\text{true}}=0$. In this case the Trotter structure of the circuit is utilized. For all but $t=0.1$, two steps of the second order formula were used, each with time step $t/2$, to reach time $t$. The gate sequence of this circuit can be rearranged so that when the second step has time step $-t/2$, the first step will be undone and the resultant state will be the initial state. As such all expectation values of Pauli $Z$ strings will be $1$ and so ODR can be used at a minimal addition to circuit depth.

 To generate the points in Fig.~\SubFigRef{fig:quantum_results}{a}, between $20$ to $60$ Pauli twirls are used with eight TREX twirls per Pauli twirl and $100$ shots per circuit. The TREX twirls are averaged together and then ODR is applied to each Pauli twirl. The resultant renormalized Pauli twirls are then bootstrap resampled to get the mean and standard error for each run. These values are reported in Table~\ref{tab:reported_values}, which also includes the depth of the circuit for each time step as well as the exact value of the Trotterized evolution. The exact value is calculated by Exact Diagonalization.
 \begin{table}[h]
     \centering
     \begin{tabularx}{\columnwidth}{X X X X}
     \toprule
     Time $t$ & depth & ibm\_fez & Exact \\
     \midrule
     0.1 & 25 & 0.0637 (16) & 0.0683 \\
     0.2 & 48 & 0.2013 (41) & 0.2015 \\
     0.3 & 48 & 0.3834 (41) & 0.3846 \\
     0.4 & 51 & 0.5509 (54) & 0.5510 \\
     0.5 & 51 & 0.6269 (82) & 0.6137 \\
     0.6 & 51 & 0.549 (11) & 0.5471 \\
     \bottomrule
     \end{tabularx}
     \caption{Estimated and exact values of $\langle H_B\rangle$ as a function of time}
     \label{tab:reported_values}
 \end{table}

 \section{Discussions and Conclusion}
\label{sec:Conclusions}
Gauge theories are ubiquitous in descriptions of nature, appearing in the Standard Model, many condensed matter systems and theories of gravity. $\textit{Ab initio}$ calculations of their properties are vital for comparing theoretical predictions to experimental results. This is particularly true for QCD, whose strongly-coupled nature at low energies gives rises to a complex array of emergent phenomena as well as important contributions to high-energy processes. While massive theoretical and algorithmic developments in classical computing have allowed us to probe many aspects of gauge theories, there remain a plethora of open questions that do not seem amenable to these methods. For example, the determination of the phase diagram of finite-density QCD and direct, nonperturbative calculations of real-time dynamics of hadrons appear inaccessible via classical techniques due to sign and signal-to-noise problems. Even our knowledge of whether the electroweak gauge interactions in the Standard Model can be given a precise, nonperturbatively regularized definition using lattice field theory faces similar difficulties, as sign problems obstruct the verification of candidate regularizations of chiral gauge theories. 

Quantum simulations of Hamiltonian formulations of lattice gauge theories provide a powerful alternative framework for exploring strongly correlated quantum systems. There is an increased need for theoretical tools that make target systems amenable for such quantum simulations. For lattice gauge theories, a necessary step is constructing a finite-dimensional representation of the Hamiltonian, so that the continuous gauge groups of the Standard Model can be encoded into a discrete register of qubits. Ideally, this encoding can be efficiently simulated at all values of the (bare) gauge coupling and is also systematically improvable. In the last decade, many Hamiltonians for both Abelian and non-Abelian gauge theories have been constructed. The proposal put forth in Refs.~\cite{DAndrea:2023qnr,Grabowska:2024emw}, combining  gauge fixing and a basis inspired by the axis-angle representation of \sutwo, is one such promising approach for simulating real-time dynamics that have hitherto been inaccessible on classical devices.

The focus of this work was constructing the full algorithmic pipeline for simulating the real-time dynamics of the fully gauge-fixed two plaquette system with open boundary conditions, working in the mixed basis. While the explicit focus was on the two plaquette system, many of the developments are applicable to larger systems in two- and three-dimensions. The starting point was from the Hamiltonian written in the group element basis and ended with the measurement of an observable on a quantum processor. Throughout the development of this workflow, several key points were addressed.

First, this work validated that previously introduced mixed-basis truncation schemes work well for larger system sizes; an avenue for additional study is whether this scheme, particularly the optimal value of $\omega_\text{max}$ given in Eq.~\eqref{eq:OmegaMax}, will continue to provide small truncation errors as system size is increased further. An important development in this work was demonstrating how the inclusion of first-order derivatives, only present in systems with two or more plaquettes, inhibits the exponential convergence predicted by the Nyquist-Shannon theorem. In particular, approximating first derivatives with finite-differences introduces digitization errors that scale polynomially with the truncation scale. This work demonstrated that with a qubit allowance of three qubits per plaquette, the low-energy properties of the two-plaquette system could be recovered with a precision of $10^{-3}$. However, it is not clear how this precision depends on the lattice volume and therefore an important question that must be addressed in future work is whether the inclusion of first-derivative terms dramatically limits the precision achievable in this formulation. Such first derivative terms appear in many other quantum many-body systems beyond \sutwo, and developing formulations of such terms compatible with quantum simulation is an area where more developments are needed.

Second, this work presented two distinct algorithms for implementing time-evolution on a digital quantum computer. The two algorithms had different quantum resource scalings and thus had different regimes of applicability. The first, described in Sec.~\ref{subsec:DifOpCircuits}, sorted the Hamiltonian according to the highest order differential operator that each term contained. A sub-circuit was constructed for each type of term. The benefit of this algorithm is that it scales more favorably in circuit depth for a large number of qubits and gives a Trotterization scheme with less error. Additionally, it can be easily adapted for systems with a larger number of plaquettes. However, it could not be implemented on currently-available quantum processors due to its depth. The second algorithm, presented in Sec.~\ref{subsec:coarse}, decomposed the Hamiltonian directly into Pauli strings, truncating the Pauli strings that had a coefficient below some \textit{a posteriori} determined value. This resulted in a quantum circuit that was sufficiently small as to be simulated on \textbf{ibm\_fez}.

Lastly, this work determined an error mitigation strategy for simulating the real-time dynamics of the two-plaquette system on NISQ-era hardware. Using previously-developed methods, namely Dynamical Decoupling, Pauli Twirling, TREX and Operator Decoherence Renormalization, were all necessary to achieve percent-level agreement between the quantum and classical results. It is not yet clear how the necessary quantum resource will scale with lattice volume and gauge coupling. Furthermore, different low-energy observables may require different error mitigation strategies.

The fully-gauged fixed two-plaquette \sutwo Hamiltonian is a very fruitful system for understanding some of the hurdles that lie ahead on the path to simulating QCD on quantum processors. It is a relatively simple system, exactly solvable using FEM, but contains both non-Abelian gauge dynamics and non-trivial interacting degrees of freedom. This work charts a course towards simulations of larger two- and three-dimensional systems while also demonstrating the viability of the mixed-basis formulation for studying the properties of \sutwo gauge theories at all values of the gauge coupling.

\begin{acknowledgments}
We gratefully acknowledge discussion with Roland Farrell, Martin Savage, and Nikita Zemlevskiy.
This work was supported, in part, 
by U.S. Department of Energy, Office of Science, Office of Nuclear Physics, InQubator for Quantum Simulation (IQuS)\footnote{\url{https://iqus.uw.edu}} under Award Number DOE (NP) Award DE-SC0020970 via the program on Quantum Horizons: QIS Research and Innovation for Nuclear Science\footnote{\url{https://science.osti.gov/np/Research/Quantum-Information-Science}}.
It was also supported, in part, by the Department of Physics\footnote{\url{https://phys.washington.edu}}
and the College of Arts and Sciences\footnote{\url{https://www.artsci.washington.edu}} at the University of Washington. 
We acknowledge the use of IBM Quantum services for this work. The views expressed are those of the authors, and do not reflect the official policy or position of IBM or the IBM Quantum team. We have made extensive use of Wolfram {\tt Mathematica}~\cite{Mathematica},
{\tt python}~\cite{python3,Hunter:2007}, {\tt jupyter} notebooks~\cite{PER-GRA:2007} 
in the {\tt Conda} environment~\cite{anaconda},
and IBM's quantum programming environment {\tt qiskit}~\cite{qiskit}.
\end{acknowledgments}

\bibliographystyle{apsrev4-1}
\bibliography{refs}

\clearpage
\appendix

\section{Explicit Form of Quantum Circuits}
\label{sec:appendixA}
This section gives explicit forms for the circuits that are constructed in Sec.~\ref{subsec:coarse} by specifying the Pauli decomposition $\mathcal{D}(\htp)$. The circuit used to calculate Fig.~\SubFigRef{fig:quantum_results}{a} is formed by exponentiating the terms in this decomposition. Table~\ref{tab:delta_high} gives the terms in the decomposition that were retained for the circuits implemented on the quantum device, whereas Table~\ref{tab:delta_low} gives the terms in the decomposition that were dropped. 

The \textbf{subcircuit} column specifies the specific subcircuit, $U_1,U_2$ or $U_3$, represented in accordance with their graphical depiction in Fig.~\SubFigRef{fig:circuit_stuff}{b}. The \textbf{Pauli String} column denotes a specific Pauli string, where the left-most index corresponds to the top wire in the circuit element and the rightmost corresponds to the bottom wire. The $\delta$ column denotes at which truncation level the corresponding string would be dropped. For Fig.~\SubFigRef{fig:quantum_results}{a}, the truncation level of $0.66$ is chosen.  As such, all entries in Table~\ref{tab:delta_high} have values $\delta>0.66$ and all entries in Table~\ref{tab:delta_low} have values $\delta<0.66$.

\section{Explicit Matrix Elements of Hamiltonian Operators}
This appendix contains the explicit expression for various electric bilinear operators that appear in the Hamiltonian. The matrix elements of these operators are presented in three different bases: group element, mixed and character irrep. 

\subsection{Group Element Basis}
\label{app:GEB}
In the group element basis, the Hilbert space basis states are spanned by the three continuous variables, $\omega, \theta, \phi$. Using this representation, the electric bilinear matrix elements are given by
\begin{align}\label{eq:electric_terms}
\CE_1^2&= - \left(\pdv[2]{}{\omega_1}+ \cot \frac{\omega_1}{2}\pdv{}{\omega_1}\right)+\frac{1}{4}\csc^2\frac{\omega_1}{2} \CN \nonumber \\
\CE_2^2&= - \left(\pdv[2]{}{\omega_2}+ \cot \frac{\omega_2}{2}\pdv{}{\omega_2}\right)+\frac{1}{4}\csc^2\frac{\omega_2}{2} \CN \nonumber \\
\boldsymbol{{\cal E}}_{R2}\cdot \boldsymbol{{\cal E}}_{L2} &= \boldsymbol{\CE}_2^2-\frac{1}{2}\CN \nonumber \\
\boldsymbol\CE_{1R}\cdot \boldsymbol \CE_{2R} &=-\cos \Theta \pdv{}{\omega_1}{ \omega_2}\nonumber\\
&+\frac{1}{2}\sin \Theta\left(\cot \frac{\omega_2}{2}\pdv{}{\omega_1}+\cot \frac{\omega_1}{2}\pdv{}{\omega_2}\right.\nonumber\\
&+\left.\frac{1}{2}\cot \frac{\omega_1}{2}\cot \frac{\omega_2}{2}\pdv{}{\Theta}\right) \nonumber \\
&-\frac{1}{4}\left(1+\cos \Theta \cot \frac{\omega_1}{2}\cot \frac{\omega_2}{2} \right) \CN \nonumber \\
\boldsymbol\CE_{1R}\cdot \boldsymbol \CE_{2L}&= \boldsymbol\CE_{1R}\cdot \boldsymbol \CE_{2R}+\frac{1}{2}\CN
\end{align}
where $\CN$ is the operator for which the Legendre polynomials are eigenfunctions, given in Eq.~\eqref{eq:LegDifEq}

\subsection{Mixed Basis}
\label{app:MB}
The mixed basis and the group element basis are related via
\begin{align}
\braket{\omega_1' \, \omega_2'\, \nu}{\omega_1 \, \omega_2\, \Theta} = P_\nu (\Theta)\delta(\omega_1-\omega_1')\delta(\omega_2-\omega_2')
\end{align}
To convert the Hamiltonian from the group element basis to the mixed basis, these four integrals
\begin{align}
\int d(\cos\Theta)\, P_{\nu'}(\Theta)P_{\nu}(\Theta)&= \delta_{\nu'\nu} \\
\int d(\cos\Theta)\,P_{\nu'}(\Theta)\CN P_{\nu}(\Theta)&= \nu(\nu+1)\delta_{\nu'\nu} \nonumber \\
\int d(\cos\Theta)\, P_{\nu'}(\Theta)\cos \Theta P_{\nu}(\Theta)&=\frac{(\nu+1)\, \delta_{\nu'\nu+1}}{\sqrt{(2\nu+1)(2\nu+3)}} \nonumber\\
&+ \frac{\nu\, \delta_{\nu'\nu-1}}{\sqrt{(2\nu+1)(2\nu-1)}}\nonumber \\
\int d(\cos\Theta) P_{\nu'}(\Theta)\sin \Theta \pdv{P_{\nu}(\Theta)}{\Theta}&= \frac{\nu\left(\nu+1\right) \,\delta_{\nu'\nu+1}}{\sqrt{(2\nu+1)(2\nu+3)}}\nonumber\\
&-\frac{\nu\left(\nu+1\right)\,\delta_{\nu'\nu-1}}{\sqrt{(2\nu+1)(2\nu-1)}} \nonumber
\end{align}
are used. Note that the unconventional normalization on $P_\nu(\Theta)$ is due to the requirement that the Hilbert space basis be orthonormal.

\begin{table*}[t]
\centering
\renewcommand{\arraystretch}{1.2}

\begin{tabular}{|c||c|c|c|}
\hline
\textbf{Subcircuit}
& \textbf{Pauli String}
&\textbf{Coefficient}
&\textbf{Truncation Percentage} $\boldsymbol{\delta}$\\
\hline

\multirow[t]{3}{*}{%
\parbox[t]{0.3\textwidth}{%
\vspace{-55pt}
\centering
\includegraphics[width=0.3\textwidth]{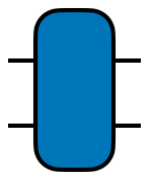}
}}
&&&\\
& I$\,\otimes\,$X & -4.358185 & 0.9857 \\
& Z$\,\otimes\,$Z & 3.049102 & 0.9429 \\
& X$\,\otimes\,$X & -2.468268 & 0.9143 \\
& X$\,\otimes\,$I & 1.565659 & 0.8571 \\
& Y$\,\otimes\,$Y & -1.396263 & 0.7857 \\
& I$\,\otimes\,$Z & 0.884597 & 0.7429 \\
& Z$\,\otimes\,$I & 0.519954 & 0.7143 \\
&&&\\
\hline

\multirow[t]{3}{*}{%
\parbox[t]{0.3\textwidth}{%
\vspace{-70pt}
\centering
\includegraphics[width=0.3\textwidth]{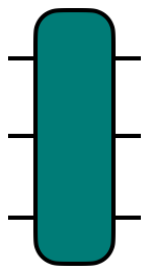}
}}
&&&\\
& Z$\,\otimes\,$I$\,\otimes\,$I & -3.778734 & 0.9571 \\
& Z$\,\otimes\,$Z$\,\otimes\,$I & -1.678690 & 0.8857 \\
& Z$\,\otimes\,$I$\,\otimes\,$Z & -1.452504 & 0.8143 \\
& Z$\,\otimes\,$Z$\,\otimes\,$Z & -1.392269 & 0.7571 \\
&&&\\
\hline

\multirow[t]{3}{*}{%
\parbox[t]{0.3\textwidth}{%
\vspace{-70pt}
\centering
\includegraphics[width=0.3\textwidth]{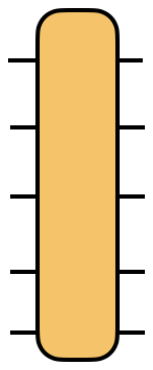}
}}
&&&\\
&&&\\
& X$\,\otimes\,$I$\,\otimes\,$I$\,\otimes\,$I$\,\otimes\,$I & 0.283820 & 0.6857 \\
& Y$\,\otimes\,$I$\,\otimes\,$Y$\,\otimes\,$I$\,\otimes\,$I & -0.226330 & 0.6714 \\
& & & \\
&&&\\
\hline
\end{tabular}
\caption{Terms in the Pauli string decomposition of $\htp$ that are retained in the circuits. The colored circuit elements correspond to those in the main text. The Pauli string column denote which the rotations $e^{-ic_p P}$ in the circuit for the string $P$ and corresponding coefficient $c_p$. The truncation percentage $\delta$ denotes the fraction of Pauli strings that would need to be truncated for this term to have the smallest coefficient. As an example, all terms in this table were included in the circuit which was run at $\delta=.66$ and so all $\delta$ values in this table are greater than $.66$.}
\label{tab:delta_high}
\end{table*}

\begin{table*}[t]
\centering
\renewcommand{\arraystretch}{1.2}

\begin{tabular}{|c||c|c|c|}
\hline
\textbf{Subcircuit}
& \textbf{Pauli String}
& \textbf{Coefficient}
& \textbf{Truncation Percentage} $\delta$ \\
\hline

\multirow[t]{45}{*}{%
\parbox[t]{0.3\textwidth}{%
\vspace{200pt}
\centering
\includegraphics[width=0.3\textwidth]{subcirc3.pdf}
}}
&&&\\
& Y$\,\otimes\,$I$\,\otimes\,$I$\,\otimes\,$I$\,\otimes\,$Y & -0.226330 & 0.6571 \\
& X$\,\otimes\,$I$\,\otimes\,$I$\,\otimes\,$Z$\,\otimes\,$I & 0.189503 & 0.6429 \\
& X$\,\otimes\,$Z$\,\otimes\,$I$\,\otimes\,$I$\,\otimes\,$I & 0.189503 & 0.6286 \\
& X$\,\otimes\,$I$\,\otimes\,$Y$\,\otimes\,$I$\,\otimes\,$Y & -0.180485 & 0.6143 \\
& Y$\,\otimes\,$I$\,\otimes\,$Y$\,\otimes\,$Z$\,\otimes\,$I & -0.151118 & 0.6000 \\
& Y$\,\otimes\,$Z$\,\otimes\,$I$\,\otimes\,$I$\,\otimes\,$Y & -0.151118 & 0.5857 \\
& X$\,\otimes\,$I$\,\otimes\,$I$\,\otimes\,$I$\,\otimes\,$Z & 0.136188 & 0.5714 \\
& X$\,\otimes\,$I$\,\otimes\,$Z$\,\otimes\,$I$\,\otimes\,$I & 0.136188 & 0.5571 \\
& X$\,\otimes\,$Z$\,\otimes\,$I$\,\otimes\,$Z$\,\otimes\,$I & 0.126529 & 0.5429 \\
& Y$\,\otimes\,$X$\,\otimes\,$Y$\,\otimes\,$I$\,\otimes\,$I & 0.113165 & 0.5286 \\
& Y$\,\otimes\,$Y$\,\otimes\,$X$\,\otimes\,$I$\,\otimes\,$I & -0.113165 & 0.5143 \\
& Y$\,\otimes\,$I$\,\otimes\,$I$\,\otimes\,$X$\,\otimes\,$Y & 0.113165 & 0.5000 \\
& Y$\,\otimes\,$I$\,\otimes\,$I$\,\otimes\,$Y$\,\otimes\,$X & -0.113165 & 0.4857 \\
& X$\,\otimes\,$I$\,\otimes\,$I$\,\otimes\,$Z$\,\otimes\,$Z & 0.110025 & 0.4714 \\
& X$\,\otimes\,$Z$\,\otimes\,$Z$\,\otimes\,$I$\,\otimes\,$I & 0.110025 & 0.4571 \\
& Y$\,\otimes\,$I$\,\otimes\,$Y$\,\otimes\,$I$\,\otimes\,$Z & -0.108602 & 0.4429 \\
& Y$\,\otimes\,$I$\,\otimes\,$Z$\,\otimes\,$I$\,\otimes\,$Y & -0.108602 & 0.4286 \\
& X$\,\otimes\,$I$\,\otimes\,$Z$\,\otimes\,$Z$\,\otimes\,$I & 0.090931 & 0.4143 \\
& X$\,\otimes\,$Z$\,\otimes\,$I$\,\otimes\,$I$\,\otimes\,$Z & 0.090931 & 0.4000 \\
& X$\,\otimes\,$I$\,\otimes\,$Y$\,\otimes\,$X$\,\otimes\,$Y & 0.090243 & 0.3857 \\
& X$\,\otimes\,$I$\,\otimes\,$Y$\,\otimes\,$Y$\,\otimes\,$X & -0.090243 & 0.3714 \\
& X$\,\otimes\,$X$\,\otimes\,$Y$\,\otimes\,$I$\,\otimes\,$Y & 0.090243 & 0.3571 \\
& X$\,\otimes\,$Y$\,\otimes\,$X$\,\otimes\,$I$\,\otimes\,$Y & -0.090243 & 0.3429 \\
& Y$\,\otimes\,$I$\,\otimes\,$Y$\,\otimes\,$Z$\,\otimes\,$Z & -0.087739 & 0.3286 \\
& Y$\,\otimes\,$Z$\,\otimes\,$Z$\,\otimes\,$I$\,\otimes\,$Y & -0.087739 & 0.3143 \\
& Y$\,\otimes\,$X$\,\otimes\,$Y$\,\otimes\,$Z$\,\otimes\,$I & 0.075559 & 0.3000 \\
& Y$\,\otimes\,$Y$\,\otimes\,$X$\,\otimes\,$Z$\,\otimes\,$I & -0.075559 & 0.2857 \\
& Y$\,\otimes\,$Z$\,\otimes\,$I$\,\otimes\,$X$\,\otimes\,$Y & 0.075559 & 0.2714 \\
& Y$\,\otimes\,$Z$\,\otimes\,$I$\,\otimes\,$Y$\,\otimes\,$X & -0.075559 & 0.2571 \\
& X$\,\otimes\,$Z$\,\otimes\,$I$\,\otimes\,$Z$\,\otimes\,$Z & 0.073463 & 0.2429 \\
& X$\,\otimes\,$Z$\,\otimes\,$Z$\,\otimes\,$Z$\,\otimes\,$I & 0.073463 & 0.2286 \\
& X$\,\otimes\,$I$\,\otimes\,$Z$\,\otimes\,$I$\,\otimes\,$Z & 0.065348 & 0.2143 \\
& Y$\,\otimes\,$I$\,\otimes\,$Z$\,\otimes\,$X$\,\otimes\,$Y & 0.054301 & 0.2000 \\
& Y$\,\otimes\,$I$\,\otimes\,$Z$\,\otimes\,$Y$\,\otimes\,$X & -0.054301 & 0.1857 \\
& Y$\,\otimes\,$X$\,\otimes\,$Y$\,\otimes\,$I$\,\otimes\,$Z & 0.054301 & 0.1714 \\
& Y$\,\otimes\,$Y$\,\otimes\,$X$\,\otimes\,$I$\,\otimes\,$Z & -0.054301 & 0.1571 \\
& X$\,\otimes\,$I$\,\otimes\,$Z$\,\otimes\,$Z$\,\otimes\,$Z & 0.052795 & 0.1429 \\
& X$\,\otimes\,$Z$\,\otimes\,$Z$\,\otimes\,$I$\,\otimes\,$Z & 0.052795 & 0.1286 \\
& X$\,\otimes\,$X$\,\otimes\,$Y$\,\otimes\,$X$\,\otimes\,$Y & -0.045121 & 0.1143 \\
& X$\,\otimes\,$X$\,\otimes\,$Y$\,\otimes\,$Y$\,\otimes\,$X & 0.045121 & 0.1000 \\
& X$\,\otimes\,$Y$\,\otimes\,$X$\,\otimes\,$X$\,\otimes\,$Y & 0.045121 & 0.0857 \\
& X$\,\otimes\,$Y$\,\otimes\,$X$\,\otimes\,$Y$\,\otimes\,$X & -0.045121 & 0.0714 \\
& Y$\,\otimes\,$X$\,\otimes\,$Y$\,\otimes\,$Z$\,\otimes\,$Z & 0.043870 & 0.0571 \\
& Y$\,\otimes\,$Y$\,\otimes\,$X$\,\otimes\,$Z$\,\otimes\,$Z & -0.043870 & 0.0429 \\
& Y$\,\otimes\,$Z$\,\otimes\,$Z$\,\otimes\,$X$\,\otimes\,$Y & 0.043870 & 0.0286 \\
& Y$\,\otimes\,$Z$\,\otimes\,$Z$\,\otimes\,$Y$\,\otimes\,$X & -0.043870 & 0.0143 \\
& X$\,\otimes\,$Z$\,\otimes\,$Z$\,\otimes\,$Z$\,\otimes\,$Z & 0.042652 & 0.0000 \\
&&&\\
\hline
\end{tabular}
\caption{The columns are the same as in the Table~\ref{tab:delta_high}. This table contains all of the Pauli string in the decomposition $\mathcal{D}(\htp)$ that were dropped in the quantum simulations. As such all entries in the $\delta$ column are less than $\delta=.66$. Note that all of the terms dropped are included in the yellow circuit element. They all have either an $X$ or a $Y$ located on the $\nu$ register. Additionally, they are generally much higher weight and have support on a greater number of qubits.}
\label{tab:delta_low}
\end{table*}

\subsection{Character Irrep Basis}
\label{app:CIB}
The characer irrep basis and the group element basis are related via
\begin{align}
\braket{J_1 J_2 \nu}{\omega_1\omega_2\Theta} = \chi^{J_1}_\nu(\omega_1) \chi^{J_2}_\nu (\omega_2) P_\nu (\Theta)
\end{align}
where $\chi^J_\nu(\omega)$ are generalized character functions of the irreducible representations of the rotation group. These functions are normalized such that
\begin{align}
\int_0^{2\pi} 4 \sin^2\frac{\omega}{2}\chi^{J'}_{\nu}(\omega) \chi^{J}_{\nu}(\omega) = \delta_{JJ'}
\end{align}
where the factor of four is due to the explicit form of the Haar measure for \sutwo parameterized using the axis-angle coordinates.

To convert between the group element basis and the character irrep basis, several matrix elements need to be computed. For the magnetic Hamiltonian, the only relevant matrix element is
\begin{align}
&\mel{J'_1\,J'_2\,\nu'}{\Tr \hat X_1}{J_1\,J_2\,\nu}= \frac{1}{ \sqrt{2}}\,\delta_{J_2'J^{\phantom{'}}_2}\delta_{\nu'\nu}  \\
&\hspace{15pt}\times \left(\sqrt{\frac{(2J_1-\nu +1) (2 J_1+\nu +2)}{(J_1+1) (2 J_1+1)}}\,\delta_{J_1' J^{\phantom{'}}_1+\frac{1}{2}} \right. \nonumber \\
&\hspace{15pt}\left.+\sqrt{\frac{(2 J_1-\nu ) (2J_1+\nu +1)}{J_1 (2 J_1+1)}}\,\delta_{J_1' J^{\phantom{'}}_1-\frac{1}{2}}\right) \, ,
\end{align}
with the matrix element for $\Tr \hat X_2$ the same, except with the exchange of $1\leftrightarrow 2$. For the electric component of the Hamiltonian, the relevant matrix elements are
\begin{align}
&\mel{J'_1\,J'_2\,\nu'}{\hat\CE_\kappa^2}{J_1\,J_2\,\nu}= J_\kappa(J_\kappa+1)\delta_{J_1'J^{\phantom{'}}_1}\delta_{J_2'J^{\phantom{'}}_2}\delta_{\nu'\nu}\nonumber \\
&\mel{J'_1\,J'_2\,\nu'}{\hat{\boldsymbol{\CE}}_{R2}\cdot \hat{\boldsymbol{\CE}}_{R2}}{J_1\,J_2\,\nu}=\delta_{J_1'J^{\phantom{'}}_1}\delta_{J_2'J^{\phantom{'}}_2}\delta_{\nu'\nu}\nonumber \\
&\hspace{30pt}\times\left(J_2(J_2+1)-\frac{\nu(\nu+1}{2}\right) \nonumber \\
&\mel{J'_1\,J'_2\,\nu'}{\hat{\boldsymbol{\CE}}_{R1}\cdot \hat{\boldsymbol{\CE}}_{R2}} {J_1\,J_2\,\nu}=-\frac{1}{4}\,\delta_{J_1'J^{\phantom{'}}_1}\delta_{J_2'J^{\phantom{'}}_2}\nonumber \\
&\hspace{30pt}\times\left(\nu(\nu+1)\delta_{\nu'\nu}+\delta_{\nu'\nu+1}\frac{c^+_{J_1 \nu}c^+_{J_2 \nu}(\nu+1)}{\sqrt{(2\nu+1)(2\nu+3)}}\right.\nonumber \\
&\hspace{30pt}\left.+\delta_{\nu'\nu-1}\frac{c^-_{J_1 \nu}c^-_{J_2 \nu}\,\nu }{\sqrt{(2\nu+1)(2\nu-1)}}\right) \nonumber \\
&\mel{J'_1\,J'_2\,\nu'}{\hat{\boldsymbol{\CE}}_{R1}\cdot \hat{\boldsymbol{\CE}}_{L2}}{J_1\,J_2\,\nu}=\frac{1}{4}\,\delta_{J_1'J^{\phantom{'}}_1}\delta_{J_2'J^{\phantom{'}}_2}\nonumber \\
&\hspace{30pt}\times\left(\nu(\nu+1)\delta_{\nu'\nu}-\delta_{\nu'\nu+1}\frac{c^+_{J_1 \nu}c^+_{J_2 \nu}(\nu+1)}{\sqrt{(2\nu+1)(2\nu+3)}}\right.\nonumber \\
&\hspace{30pt}\left.-\delta_{\nu'\nu-1}\frac{c^-_{J_1 \nu}c^-_{J_2 \nu}\,\nu }{\sqrt{(2\nu+1)(2\nu-1)}}\right)
\end{align}
where in the first expression $\kappa = 1,2$ and the coefficients $c^\pm_{J\nu}$ are given by
\begin{align}
c^+_{J \nu}&= \sqrt{(2 J+\nu +2) (2 J-\nu)} \nonumber \\
c^-_{J \nu}&= \sqrt{(2 J+\nu+1 ) (2 J-\nu+1)}
\end{align}

The full Hamiltonian will not be given here, though key components will. Of particular interest are the components that contain only first derivatives with respect to the $\omega$ variables, grouped into $H_\partial$. This Hamiltonian is given by
\begin{equation}
    H_{\partial} = \sum_{\nu}\ket{\nu}\bra{\nu+1}H^{+}_{\nu}+\sum_{\nu}H^{-}_{\nu}\ket{\nu}\bra{\nu-1}
\end{equation}
where $H^\pm$ are 
\begin{align}
H^{+}_{\nu} &= A_{\nu}\pdv{}{\omega_1}\pdv{}{\omega_2}+B_{\nu}\left(\cot \frac{\omega_1}{2}\pdv{}{\omega_2}+\cot \frac{\omega_2}{2}\pdv{}{\omega_1}\right)\nonumber\\
&+\frac{1}{4}\left(\nu(\nu+1)A_{\nu}-2B_{\nu}\right)\cot \frac{\omega_1}{2}\cot \frac{\omega_2}{2}\nonumber\\
H^{-}_{\nu} &= D_{\nu}\pdv{}{\omega_1}\pdv{}{\omega_2}+C_{\nu}\left(\cot \frac{\omega_1}{2}\pdv{}{\omega_2}+\cot \frac{\omega_2}{2}\pdv{}{\omega_1}\right)\nonumber\\
&+\frac{1}{4}\left(\nu(\nu+1)D_{\nu}-2C_{\nu}\right)\cot \frac{\omega_1}{2}\cot \frac{\omega_2}{2}
\end{align}
with the coefficients 
\begin{align}
    A_{\nu} &= \frac{(\nu+1)}{\sqrt{(2\nu+1)(2\nu+3)}}\nonumber\\
    B_{\nu} &= -\frac{(\nu+1)^2}{2\sqrt{(2\nu+1)(2\nu+3)}}\nonumber\\
    C_{\nu} &= \frac{\nu^2}{2\sqrt{(2\nu-1)(2\nu+1)}}\nonumber\\
    D_{\nu} &= \frac{\nu}{\sqrt{(2\nu-1)(2\nu+1)}} \, .
\end{align}
The relation that $H_{\nu}^{+}=(H_{\nu+1}^-)^{\dagger}$ ensures hermiticity of the overall expression.

For the quantum simulations presented in this work, only $H_0^+$ and $H_1^-$ is needed and so they are written out explicitly here:
\begin{align}
H^+_0&=\frac{1}{\sqrt{3}}\pdv{}{\omega_1}\pdv{}{\omega_2}+\frac{1}{4\sqrt{3}}\cot \frac{\omega_1}{2}\cot \frac{\omega_2}{2}\nonumber \\
&-\frac{1}{2\sqrt{3}}\left(\cot \frac{\omega_1}{2}\pdv{}{\omega_2}+\cot \frac{\omega_2}{2}\pdv{}{\omega_1}\right) \nonumber \\
&= (H^-_1)^\dagger \, \, .
\end{align}

\section{Table of Lowest Energy Eigenstates in the character Irrep Basis}
The energy spectrum of the two plaquette Hamiltonian in the strong coupling limit can be found exactly when working in the character irrep basis. This is due to the electric component of the  Hamiltonian being block diagonal, with each block of finite dimension $2\, \text{Min}(J_1, J_2)+1$; the magnetic component is negligible and can be ignored. The energies and expressions for the several lowest lying eigenstates are showin in Table~\ref{tab:CharIrrepStrong}.
\begin{table*}[t]
    \centering
\begin{align}
\begin{array}{|l|c|c|c|} \hline
\text{\textbf{Level of State}}& \text{\textbf{Degeneracy}} & \text{\textbf{In Irrep Basis:}}\ket{J_1, J_2, \nu} & \text{\textbf{Energy}}\\\hline
 &&&\\
\text{Ground} & \text{None} & \ket{0,\,0,\,0}& 0 \\
 &&&\\ \hline
&&&\\
&  & \ket{\frac{1}{2},\,0,\,0} &  \\
\text{First Excited}&\text{Two-Fold}&&\frac{3}{2}g^2\\
 &  & \ket{0,\,\frac{1}{2},\,0} &\\
 &&&\\ \hline
 &&&\\
\text{Second Excited} & \text{None} & \frac{1}{2}\ket{\frac{1}{2},\,\frac{1}{2},\,0}-\frac{\sqrt{3}}{2}\ket{\frac{1}{2},\,\frac{1}{2},\,1} & \frac{9}{4}g^2 \\
&&&\\ \hline
&&&\\
\text{Third Excited} & \text{None} & \frac{\sqrt{3}}{2}\ket{\frac{1}{2},\,\frac{1}{2},\,0}+\frac{1}{2}\ket{\frac{1}{2},\,\frac{1}{2},\,1} & \frac{13}{4}g^2 \\
&&&\\ \hline
&&&\\
& & \ket{1,\,0,\,0} &  \\
\text{Fourth Excited} & \text{Two-Fold}&&4g^2\\
 &  & \ket{0,\,1,\,0} & \\
 &&&\\ \hline
 &&&\\
& &\frac{1}{\sqrt{3}} \ket{1,\,\tfrac{1}{2},\,0}-\sqrt{\frac{2}{3}} \ket{1,\,\tfrac{1}{2},\,1} &  \\
\text{Fifth Excited} & \text{Two-Fold}&&\frac{9}{2}g^2\\
&&\frac{1}{\sqrt{3}} \ket{\tfrac{1}{2},\,1,\,0}-\sqrt{\frac{2}{3}} \ket{\tfrac{1}{2},\,1,\,1}&\\
 &&&\\ \hline
&&&\\
& &\frac{1}{\sqrt{3}} \ket{1,\,\tfrac{1}{2},\,0}-\sqrt{\frac{2}{3}} \ket{1,\,\tfrac{1}{2},\,1} &  \\
&&&\\
\text{Sixth Excited} & \text{Three-Fold}& \sqrt{\frac{2}{3}} \ket{\tfrac{1}{2},\,1,\,0}+ \frac{1}{\sqrt{3}}\ket{\tfrac{1}{2},\,1,\,1}&6g^2\\
 &&&\\
 &&\frac{1}{3}\ket{1,\,1,\,0}-\frac{1}{\sqrt{3}}\ket{1,\, 1,\, 1}+\frac{\sqrt{5}}{3}\ket{1,\,1,\,2}&\\
 &&&\\ \hline
\end{array} \nonumber
\end{align}
\caption{Energies and explicit eigenstate constructions for the seven lowest-lying eigenstates of the two plaquette Hamiltonian in the strong coupling limit. For the states that are degenerate in this limit, the magnetic Hamiltonian breaks the degeneracy and splits the states.}
\label{tab:CharIrrepStrong}
\end{table*}

\end{document}